\documentclass[twocolumn,
               prd,
               aps,
               superscriptaddress,
               tightenlines,
               nofootinbib,
               eqsecnum,
               amsfonts,
               amsmath,
               longbibliography]{revtex4-1}
\usepackage[flushleft]{threeparttable}
\usepackage{epsfig}
\usepackage{graphics}
\usepackage{graphicx} \usepackage[dvipsnames,table]{xcolor}
\usepackage{bm}
\usepackage{txfonts}
\usepackage{natbib}
\usepackage{amssymb}
\usepackage{xspace}

\usepackage[normalem]{ulem} 
\newcommand\redsout{\bgroup\markoverwith{\textcolor{red}{\rule[0.5ex]{8pt}{1pt}}}\ULon} 

\usepackage[colorlinks]{hyperref}
\usepackage[caption=false]{subfig}
\usepackage{url}
\usepackage{float}
\usepackage[bottom]{footmisc}
\usepackage{lineno}
\usepackage{mathrsfs}
\usepackage{makecell}
\usepackage{microtype}

\AtBeginDocument{%
    \newwrite\bibnotes
    \def\bibnotesext{Notes.bib}
    \immediate\openout\bibnotes=\jobname\bibnotesext
    \immediate\write\bibnotes{@CONTROL{REVTEX41Control}}
    \immediate\write\bibnotes{@CONTROL{%
    apsrev41Control,author="08",editor="1",pages="1",title="0",year="1"}}
    \if@filesw
    \immediate\write\@auxout{\string\citation{apsrev41Control}}%
    \fi
}

\definecolor{rred}{RGB}{218,41,28}
\hypersetup{linkcolor=NavyBlue}
\hypersetup{citecolor=NavyBlue}
\hypersetup{urlcolor=NavyBlue}

\usepackage{perpage}
\MakePerPage{footnote}

\newcommand{\Mf}{M_{\rm f}}

\newcommand{\Mo}{M_{\odot}}

\newcommand{\pSEOB}{\texttt{pSEOBNR}}
\newcommand{\SEOB}{\texttt{SEOBNR}}
\newcommand{\gm}{\mathfrak{m}}
\newcommand{\msun}{~{\rm M}_{\odot}}

\newcommand{\dd}{{\rm d}}

\newcommand{\dV}{{\rm d}^{4}x \, \sqrt{-g} \,}

\newcommand{\lame}{\lambda_{\rm e}}
\newcommand{\lamo}{\lambda_{\rm o}}

\usepackage{xcolor}

\newcommand{\AEI}{\affiliation{Max Planck Institute for Gravitational Physics (Albert Einstein Institute), Am M\"uhlenberg 1, Potsdam D-14476, Germany}}
\newcommand{\UMD}{\affiliation{Department of Physics, University of Maryland, College Park, Maryland 20742, USA}}

\begin{document}

\title{Black-hole ringdown as a probe of higher-curvature gravity theories}

\author{Hector O. Silva}     \AEI
\author{Abhirup Ghosh}       \AEI
\author{Alessandra Buonanno} \AEI \UMD

\date{\today}

\begin{abstract}
Detecting gravitational waves from coalescing compact binaries allows
us to explore the dynamical, nonlinear regime of general relativity
and constrain modifications to it.
Some of the gravitational-wave events observed by the LIGO-Virgo Collaboration have sufficiently high
signal-to-noise ratio in the merger, allowing us to probe the relaxation of the remnant black
hole to its final, stationary state --- the so-called black-hole ringdown, which is
characterized by a set of quasinormal modes.
Can we use the ringdown to constrain deviations from general relativity,
as predicted by several of its contenders?
Here, we  address this question by using an inspiral-merger-ringdown waveform
model in the effective-one-body formalism, augmented with a parametrization of
the ringdown based on an expansion in the final black hole's spin.
We give a prescription on how to include in this waveform model, the quasinormal mode frequencies
calculated on a theory-by-theory basis. In particular, we focus on theories that modify
general relativity by higher-order curvature corrections, namely,
Einstein-dilaton-Gauss-Bonnet, dynamical Chern-Simons theories, and cubic- and quartic-order
effective-field-theories of general relativity.
%
We use this parametrized waveform model to measure the ringdown properties of
the two loudest ringdown signals observed so far, GW150914 and GW200129.
We find that while the Einstein-dilaton-Gauss-Bonnet theory cannot be
constrained with these events, we can place upper bounds on the
fundamental lengthscale of cubic- ($\ell_{\rm cEFT} \leqslant 38.2$~km) and quartic-order
($\ell_{\rm qEFT} \leqslant 51.3$~km) effective-field-theories of general
relativity, and of dynamical Chern-Simons gravity ($\ell_{\rm dCS} \leqslant 38.7$~km).
The latter result is a concrete example of a theory presently unconstrained by
inspiral-only analyses which, however, can be constrained by merger-ringdown
studies with current gravitational-wave data.
\end{abstract}

\maketitle

\section{Introduction}
\label{sec:intro}

Since the first detection of gravitational waves (GWs) from a binary black-hole (BBH) merger in 2015~\cite{LIGOScientific:2016aoc},
the LIGO~\cite{LIGOScientific:2014pky} and Virgo~\cite{VIRGO:2014yos} detectors have observed about 90 GW events~\cite{LIGOScientific:2021djp} from mergers
of BHs, neutron stars (NSs) \cite{TheLIGOScientific:2017qsa,LIGOScientific:2018cki,LIGOScientific:2020aai} and their mixture~\cite{LIGOScientific:2021qlt}. These results have been confirmed
by independent analyses, which have also identified a few new GW signals~\cite{Nitz:2018imz,Nitz:2019hdf,Venumadhav:2019lyq,Zackay:2019btq,Nitz:2021zwj,Olsen:2022pin}.

The large number of GW observations has allowed us to infer relevant astrophysical~\cite{LIGOScientific:2021aug} and cosmological~\cite{LIGOScientific:2021psn} information on the compact-object population in our local Universe, and also to probe general relativity (GR) in the high-velocity, dynamical and strong-field regime of gravity~\cite{Yunes:2016jcc,LIGOScientific:2021sio}. The latter complement tests of GR in the low-velocity, quasistatic or linear regimes available with Solar-System experiments~\cite{Will:2014kxa}, binary-pulsar~\cite{Wex:2014nva,Kramer:2021jcw} and galactic-center~\cite{GRAVITY:2018ofz,Do:2019txf,EventHorizonTelescope:2019ths} observations,
and cosmological measurements~\cite{Clifton:2011jh}.

The coalescence of two BHs in GR is characterized by a long \textit{inspiral} stage, during which the
holes adiabatically and steadily come closer and closer to each other, losing energy because of the emission
of GWs. Then, they \textit{merge}, forming a common apparent horizon. Subsequently, during the \textit{ringdown} stage, the newly formed remnant object settles down to a Kerr BH emitting quasinormal modes (QNMs)~\cite{Vishveshwara:1970cc,Press:1971wr,Chandrasekhar:1975zza}. Because of the no-hair conjecture in GR~\cite{Carter:1971zc,Israel:1967wq,Hawking:1971vc,Robinson:1975bv}, the QNM (complex) frequencies of (electrically neutral) astrophysical BHs are only described by the BH's mass and spin. In GR, the QNM frequencies are labeled by the harmonic indices $(\ell,m)$ and the overtone number $n$.

Several null tests have been proposed to probe the nature of gravity with GW
signals~\cite{TheLIGOScientific:2016src,Yunes:2016jcc,Ghosh:2016xx,Ghosh:2017gfp,Abbott:2018lct,LIGOScientific:2019fpa,Abbott:2020jks,LIGOScientific:2021sio}.
They include tests of GW
generation~\cite{Arun:2006yw,Yunes:2009ke,Li:2011cg,Agathos:2013upa,Mehta:2022pcn},
where deviations in the post-Newtonian (PN) coefficients in the inspiral, and
phenomenological coefficients in the plunge and merger stages can be bounded;
tests of GW propagation~\cite{Will:1997bb,Mirshekari:2011yq}, which allow us to
set upper limits on coefficients entering generalized dispersion relations,
including the Compton wavelength associated to the mass of the graviton;
tests of the polarization of gravitational
radiation~\cite{Eardley:1973zuo,Will:2014kxa,Isi:2017fbj,LIGOScientific:2017ycc,LIGOScientific:2018dkp,Wong:2021cmp}, for which more than
two GW detectors are needed to set statistically significant bounds, and tests of the remnant properties~\cite{Meidam:2014jpa,Carullo:2018sfu,Brito:2018rfr,Carullo:2019flw,Isi:2019aib,Ghosh:2021mrv,Carullo:2021dui} in the postmerger stage. So far, none of these null tests have reported any deviation from GR.

Probing the gravitational properties of the remnant object during the
ringdown, has attracted a lot of attention in the last twenty years.
Reference~\cite{Dreyer:2003bv} proposed the idea of employing BH spectroscopy~\cite{Detweiler:1980gk}
of the ringdown stage to rule out (or constrain) either modified theories of
gravity or exotic compact objects (in GR) rather than BHs, thus
testing the no-hair conjecture. Since then, many studies have quantified the accuracy
with which the QNM frequencies can be measured for
GW sources detectable with ground- and space-based detectors (see, e.g., Refs.~\cite{Berti:2005ys,Baibhav:2020tma,Bhagwat:2021kwv,Ota:2021ypb}). Several analyses~\cite{Carullo:2018sfu,
Brito:2018rfr,Carullo:2019flw,Isi:2019aib,Ghosh:2021mrv,Carullo:2021dui}
have used the GW observations of the LIGO-Virgo-KAGRA (LVK) collaboration to set upper
limits on deviations in the QNM frequencies of BHs in GR.
Others have claimed
the measurements of QNMs beyond the dominant $(2,2,0)$ mode~\cite{Capano:2021etf},
or overtones --- for example the $(2,2,1)$ mode~\cite{Isi:2019aib,Isi:2022mhy,Finch:2022ynt}
, although the evidence for overtones can also be due to noise~\cite{Cotesta:2022pci}.
These studies have been pursued either using a superposition
of damped sinusoids~\cite{Carullo:2019flw,Isi:2021iql}, in some cases augmented with QNM amplitudes calibrated to
numerical-relativity (NR) simulations, or with parametrized inspiral-merger-ringdown (IMR)
waveform models, where the QNM frequencies are not necessarily fixed
to the GR values for BHs, but kept free~\cite{Brito:2018rfr,Ghosh:2021mrv}.

Here, we will employ the parametrized IMR model of Ref.~\cite{Ghosh:2021mrv}, constructed from a nonprecessing-spin effective-one-body (EOB) waveform model~\cite{Bohe:2016gbl,Cotesta:2018fcv,Mihaylov:2021bpf}, to carry out theory-specific tests of the
ringdown using four high-curvature gravity theories. Previously, such parametrized
waveform model was employed in Refs.~\cite{Abbott:2020jks,LIGOScientific:2021sio} for theory-independent tests of the ringdown.
More specifically, here we will focus on four modified gravity theories, Einstein-dilaton-Gauss-Bonnet gravity, dynamical
Chern-Simon gravity, cubic and quartic effective-field theories (EFTs) of GR, and
express the QNM frequencies using the parametrized ringdown spin-expansion coefficients (ParSpec)
of Ref.~\cite{Maselli:2019mjd}.
In this framework, the non-GR QNM frequencies are
recast as deviations from the GR QNM values, and are expressed in terms of a single free parameter,
the fundamental lengthscale $\ell_{\rm th}$ of the gravity theory under consideration, the GR limit
corresponding to $\ell_{\rm th} \rightarrow 0$.

With this formulation of the ringdown, we use the two loudest merger-ringdown GW events, so far
observed by the LVK collaboration, notably GW150914 and GW200129, and use Bayesian-inference techniques to
perform null tests. We find no indication that GR is violated and, when possible, we place upper limits, at $90\%$ credible level,
on the lengthscale $\ell_{\rm th}$ of each theory. In Table~\ref{tab:bound_summary}, we summarize our results, and compare
them with existing constraints.

\begin{table}[t]
\begin{threeparttable}
\begin{tabular}{c | c c}
\hline \hline
Theory & Constraint & This work \\
\hline
EdGB        & $\ell_{\rm EdGB} \leqslant 1.18$~km~(GW)~\cite{Lyu:2022gdr} & -- \\
dCS         & $\ell_{\rm dCS} \leqslant 8.5$~km~(EM+GW)~\cite{Silva:2020acr}  & $\ell_{\rm dCS} \leqslant 38.7$~km \\
cubic EFT   & -- & $\ell_{\rm cEFT} \leqslant 38.2$~km \\
quartic EFT & $\ell_{\rm qEFT} \leqslant 150$~km~\cite{Sennett:2019bpc}  & $\ell_{\rm qEFT} \leqslant 51.3$~km~\tnote{a}
\\
\hline \hline
\end{tabular}
\caption{Summary of the upper limits, at $90\%$ credible level,
on the lengthscale of the modified gravity theories under investigation.
They were obtained combining Bayesian-inference results from GW150914 and GW200129.
Current constraints on $\ell_{\rm th}$ are also listed.}
\begin{tablenotes}
\item[a]{\footnotesize We notice that our result for the quartic EFT of GR is only in marginal tension with our hard cutoff scale for the
validity of the theory, and hence we do still quote it here (see Secs.~\ref{sec:remarks} and~\ref{sec:results_qeft}
for details).}
\end{tablenotes}
\label{tab:bound_summary}
\end{threeparttable}
\end{table}

The paper is organized as follows. In Sec.~\ref{sec:review_theories}, we briefly describe the four
higher-curvature modified gravity theories for which we perform the ringdown test.
In Sec.~\ref{sec:method} we build our parametrized IMR model for nonprecessing-spin compact-object
binaries making use of the ParSpec framework. After reviewing the Bayesian inference method,
in Sec.~\ref{sec:pe}, we motivate our selection of GW events from the LVK catalog, and also
discuss the range of validity of our analyses. In particular, we discuss the impact on
our results of the assumptions underlying the ParSpec framework, and the fact that
our modified gravity theories have to be interpreted as EFTs. In Sec.~\ref{sec:results},
we present our results obtained by applying Bayesian analysis on the LVK data of
GW150914 and GW1200129, and discuss how we set the upper limits
on the fundamental lengthscales $\ell_{\rm th}$. Finally, in Sec.~\ref{sec:conclusions} we
summarize our findings, and discuss how to make our framework more robust, in view
of stronger GW events in upcoming GW observational runs, by including  physical effects currently absent in our
study (e.g., precessing-spins and eccentricity). In the Appendix~\ref{app:map_details}
we provide details in calculating the non-GR QNM frequencies, when using ParSpec,
for the modified gravity theories under consideration.
Henceforth, unless otherwise specified, we work in geometric units $G = 1 = c$.

\section{Overview of modified gravity theories}
\label{sec:review_theories}

We will treat the modified theories of gravity as EFTs, and focus on
finite-size effects (see, e.g., Ref.~\cite{Sennett:2019bpc}). Thus, for each gravity theory we impose that
the fundamental lengthscale $\ell_{\rm th} \lesssim GM/c^2$, where $M$ is the mass of the BH. This implies
that observable deviations from GR present in those theories arise from modifications
to the Kerr geometry of each individual BH. Those finite-size effects can manifest themselves in
the QNMs of the remnant produced by the merger, but also in the GW phasing of the inspiral through
corrections to the GR spin-induced quadrupole and Love numbers. However, here we will not consider the latter,
instead, we will only study the impact of finite-size effects on the QNMs of the
remnant.

We start by briefly reviewing the modified gravity theories we consider in this
paper, what the current observational constraints are and what we know
about BH QNMs in each of these theories.

\subsection{Einstein-dilaton-Gauss-Bonnet gravity}
\label{sec:review_edgb}

This theory belongs to the class of scalar-Gauss-Bonnet theories, which are
described by the action
\begin{align} \label{eq:action_sgb}
    S_{\rm EdGB} = \frac{1}{16 \pi}
    \int \dV
    \left[
    R - \tfrac{1}{2}(\partial \varphi)^2
    + \tfrac{1}{4} \ell^{2}_{\rm EdGB} f(\varphi) \, \mathcal{G}
    \right],
    \nonumber \\
\end{align}
where $g \equiv \textrm{det}(g_{\mu \nu})$ is the metric determinant, $R$ is the Ricci
scalar, $\varphi$ is a dynamical scalar field, with kinetic term $(\partial \varphi)^2 = g^{\mu\nu} \partial_{\mu} \varphi \partial_{\nu} \varphi$,
which couples to the Gauss-Bonnet invariant
$\mathcal{G} =
R^{\mu\nu\rho\sigma}R_{\mu\nu\rho\sigma}
- 4 R^{\mu\nu}R_{\mu\nu}
+ R^2$,
and $R_{\mu\nu\rho\sigma}$ and $R_{\mu\nu}$ are the Riemann and Ricci tensors respectively.
By itself, $\int \dV \mathcal{G}$ is a boundary term in four dimensions
and hence does not contribute to the field equations~\cite{Myers:1987yn}.
However,  when coupled to $\varphi$, it can contribute to the field equation
through the coupling function $f(\varphi)$. The strength of the coupling is
set by $\ell_{\rm GB}$, with dimensions of length.

Different subclasses of this theory are determined by the function $f(\varphi)$
and can be divided into two classes based on the properties of their BH
solutions.
In the first class, the first derivative of the coupling function $f'(\varphi)
\equiv \dd f  / \dd \varphi$ is always nonzero and BHs are known to always
support scalar hair.
This is the case of Einstein-dilaton-Gauss-Bonnet (EdGB) gravity, for which
$f(\varphi) = \exp(\varphi)$~\cite{Kanti:1995vq}.
In the second class, $f'(\varphi) = 0$ can vanish for some constant $\varphi_0$.
In this case, the theory admits the same stationary, asymptotically flat BH
solutions as GR~\cite{Silva:2017uqg} and those of scalarized
BHs~\cite{Doneva:2017bvd,Silva:2017uqg,Macedo:2019sem,Dima:2020yac,Herdeiro:2020wei,Berti:2020kgk}.
Examples include Gaussian $f(\varphi) \propto \exp(-\varphi^2)$~\cite{Doneva:2017bvd} and
the quadratic $f(\varphi) \propto \varphi^2$~\cite{Silva:2017uqg} coupling functions.

BHs in EdGB gravity have scalar hair, to which we can associate a monopole scalar charge,
related to the asymptotic $r^{-1}$ fall-off of the scalar field, where $r$ is the distance from the BH.
This charge is not an independent parameter, and depends on the BH's mass and spin,
thus being a ``secondary hair''~\cite{Coleman:1991ku,Kanti:1995vq,Herdeiro:2015waa}.
Since the scalar field is sourced by a curvature scalar, the scalar charge is larger (smaller),
the smaller (larger) the BH mass is.

These properties are not mere theoretical curiosities; they have important observational consequences.
First, the presence of the scalar charge implies that when in binaries, BHs can source
scalar-dipole radiation (see, e.g., Refs.~\cite{Yagi:2011xp,Julie:2019sab,Shiralilou:2020gah,Shiralilou:2021mfl,Julie:2022huo})
which affects the GW phase at $-1$PN order (relative to the dominant quadrupolar GR
contribution), with magnitude proportional to the difference between the
charges of binary components. This makes EdGB gravity testable with GW
observations of compact binaries where at least one component is a BH.
Indeed, Ref.~\cite{Lyu:2022gdr} placed the bound
$\ell_{\rm EdGB} \leqslant 7.1$~km
using the NSBH binaries GW200105 and
GW200115~\cite{LIGOScientific:2021qlt} while Refs.~\cite{Nair:2019iur,Perkins:2021mhb}, obtained
$\ell_{\rm EdGB} \leqslant 9.1$~km,
by stacking the posteriors of $\ell_{\rm EdGB}$ from a selection of BBHs
from the GWTC-1 and GWTC-2 catalogs~\cite{LIGOScientific:2018mvr,LIGOScientific:2020ibl}\footnote{These bounds
are, strictly speaking, valid only when the scalar field $\varphi$ is small (i.e., $\varphi \ll 1$), and
they take into account only the leading-order scalar field interaction arising from the original dilatonic coupling
(i.e., $f(\varphi) \approx \varphi$ in Eq.~\eqref{eq:action_sgb}).
This results in what is often referred to as shift-symmetric scalar-Gauss-Bonnet theory.
In this theory, NSs do not have scalar monopole charge~\cite{Yagi:2015oca}, while BHs do~\cite{Yunes:2011we,Sotiriou:2013qea,Sotiriou:2014pfa}. Finally, we note that the constant $\alpha_{\rm EdGB}$ used in Refs.~\cite{Yagi:2011xp,Nair:2019iur,Perkins:2021mhb,Lyu:2022gdr} is related to $\ell_{\rm EdGB}$ as $\ell_{\rm EdGB} = 4 \pi^{1/4} |\alpha_{\rm EdGB}|^{1/2}$.}.
Secondly, the scalar field influences the response of BHs to linearized perturbations and thus
affects the BH's QNM spectra.
The coupling between scalar field and the Gauss-Bonnet invariant,
results in a coupling between scalar perturbations and gravitational
perturbations of polar parity which, for instance, breaks the equivalence
between the QNM spectra of polar and axial gravitational perturbations~\cite{Glampedakis:2017rar,Lenzi:2021njy}
(i.e., the isospectrality~\cite{Chandrasekhar:1985kt}) of Schwarzschild BHs in GR.

Using BH perturbation theory, the QNMs for nonrotating BHs in EdGB gravity
were first computed in Refs.~\cite{Pani:2009wy,Blazquez-Salcedo:2016enn}
and were extended (in the polar sector) to leading-order in BH spin in Ref.~\cite{Pierini:2021jxd}.
See Ref.~\cite{Bryant:2021xdh} for a study in the geometrical optics limit ($\ell \gg 1$) and
Refs.~\cite{Witek:2018dmd,Okounkova:2019zep} for NR studies.

\subsection{Dynamical Chern-Simons gravity}
\label{sec:review_dcs}

This theory is described by the action~\cite{Jackiw:2003pm,Alexander:2009tp},
\begin{align} \label{eq:action_dcs}
    S_{\rm dCS} =
    \int \dV
    \left[
    \tfrac{1}{16\pi}\, R - \tfrac{1}{2}(\partial \vartheta)^2
    + \tfrac{1}{4} \ell^{2}_{\rm dCS} \, \vartheta \, {}^{*}RR
    \right],
\end{align}
where $\vartheta$ is a pseudoscalar field which couples to the Pontryagin
density
${}^{*}RR = {}^{*}R_{\mu\nu}{}^{\rho\sigma} R^{\nu\mu}{}_{\rho\sigma}$,
where ${}^{*}R^{\mu\nu\rho\sigma}$ is the dual of the Riemann tensor
defined as
${}^{*}R^{\mu\nu\rho\sigma} =
\epsilon^{\mu\nu}{}_{\gamma\delta}
R^{\gamma\delta\rho\sigma} / 2$,
and $\epsilon_{\mu\nu\gamma\delta}$ is the Levi-Civita tensor.
The variation of the Pontryagin density is a boundary
term that does not contribute to the field equations in
four-dimensions~\cite{Jackiw:2003pm}.
However, the Pontryagin density can modify the field equations when coupled to $\vartheta$;
the strength of the coupling set by $\ell_{\rm dCS}$, with dimensions of length.

The theory admits as a solution the garden-variety Schwarzschild BH of GR.
This is not the case when rotation is included and the Kerr metric is not a
solution of the theory~\cite{Jackiw:2003pm}.
These rotating BHs support a scalar field which falls off as
$r^{-2}$ asymptotically, to which we can associate a scalar dipole
charge~\cite{Yunes:2009hc,Konno:2009kg} and
the leading-order modification to the GW phase enters at 2PN~\cite{Yagi:2011xp}.
Deviations from GR at this PN order are constrained with present GW observations~\cite{LIGOScientific:2021sio}
only at the level of $\sim 50\%$ (see the constraint on the 2PN parameter $\varphi_4$ in Fig.~6 of
Ref.~\cite{LIGOScientific:2021sio}). So far, analyses that used only the inspiral portion of
the BBH GW signals were not able to set meaningful bounds on such deviation at 2PN order~\cite{Nair:2019iur,Perkins:2021mhb}. These works, as well as the analysis we do here, probe the effects of dCS in the generation of GWs.
Weak constraints of order $\sim 10^{3}$~km were obtained on this theory by considering parity-violation propagation effects in GWs~\cite{Yamada:2020zvt,Wang:2020cub,Okounkova:2021xjv}, which
show up as an amplitude birefringence between different GW polarizations~\cite{Jackiw:2003pm,Alexander:2004wk}.
Nonetheless, the theory has been constrained in Ref.~\cite{Silva:2020acr}, which found
$\ell_{\rm dCS} \leqslant 8.5$~km,
by folding data from the X-ray observations of the pulsar
PSR~J0030+0451~\cite{Lommen:2000yt,NANOGrav:2017wvv} by
NICER~\cite{Riley:2019yda,Miller:2019cac} and from the GW observation of the binary
NS GW170817~\cite{TheLIGOScientific:2017qsa,LIGOScientific:2018cki}, using
equation-of-state independent relations between NS moment of inertia and tidal deformability~\cite{Yagi:2013bca,Yagi:2013awa,Gupta:2017vsl}.

The QNMs of the Schwarzschild BH in dCS gravity were studied in
Refs.~\cite{Yunes:2007ss,Cardoso:2009pk,Molina:2010fb}, which found that scalar
perturbations couple to gravitational perturbations of axial parity, in contrast
with EdGB gravity, resulting in a breakdown of isospectrality.
The QNM spectra of slowly-rotating BHs in dCS gravity was
studied in Refs.~\cite{Srivastava:2021imr,Wagle:2021tam}.
They were also extracted from NR simulations of BH head-on collisions in Ref.~\cite{Okounkova:2019dfo}.

\subsection{Effective-field-theory of General Relativity}

Our last example of modified gravity theories are the so-called
EFTs of GR~\cite{Endlich:2017tqa,Sennett:2019bpc,Cano:2019ore,Brandhuber:2019qpg,AccettulliHuber:2020dal,deRham:2020ejn,Cano:2020cao,Cano:2021myl}.
They are described by the action
\begin{equation} \label{eq:action_eft}
    S_{\rm EFT} = \frac{1}{16 \pi}
    \int \dd^4x \sqrt{-g}
    \left[ R
    +
    \sum_{n \geqslant 2} \ell_{\rm EFT}^{2n - 2}\, L^{(2n)}
    \right]\,,
\end{equation}
where $\ell_{\rm EFT}$ is a lengthscale assumed to be small compared to the lengthscale $M$
associated with a BH (i.e., $\ell_{\rm EFT} / M \ll 1$), and $L^{(2n)}$ are corrections
that introduce higher-order curvature tensors (with $2n$ metric derivatives).

More specifically, we follow the notation of
Refs.~\cite{Cano:2020cao,Cano:2021myl} and consider up to dimension-eight
operators ($n=4$),
\begin{subequations}
\label{eq:action_mods}
\begin{align}
    L^{(6)} &= \lame R_{\mu\nu}{}^{\rho\sigma} R_{\rho\sigma}{}^{\gamma\delta} R_{\gamma \delta}{}^{\mu\nu}
    + \lamo R_{\mu\nu}{}^{\rho\sigma} R_{\rho\sigma}{}^{\gamma\delta} \tilde{R}_{\gamma \delta}{}^{\mu\nu},
    \label{eq:s_6}
    \\
    L^{(8)} &= \varepsilon_{1} \mathcal{C}^2
    + \varepsilon_{2} \tilde{\mathcal{C}}^{2}
    + \varepsilon_{3} \mathcal{C} \tilde{\mathcal{C}},
\label{eq:s_8}
\end{align}
\end{subequations}
where ${\cal C} = R_{\mu\nu\rho\sigma} R^{\mu\nu\rho\sigma}$,
$\tilde{\cal C} = R_{\mu\nu\gamma\delta} \epsilon^{\mu\nu}{}_{\rho\sigma} R^{\rho\sigma\gamma\delta}$,
and both $\lambda_{\rm o, e}$ and $\varepsilon_{i}$ (with $i=1, 2, 3$) are dimensionless parameters.
Due to the large number of free parameters in this theory, we focus on a subset of the parameter space.
In particular, we consider dimension-six and dimension-eight operators separately.
In addition, in the dimension-six case we further assume that $\lame = \lamo = 1$,
leaving us with $\ell_{\rm cEFT}$ as our single free parameter.
Similarly, in the dimension-eight case, we set $\varepsilon_{1} = 1$ and $\varepsilon_{2} = \varepsilon_{3} = 0$,
as done in Ref.~\cite{Sennett:2019bpc}.
This leaves us with $\ell_{\rm qEFT}$ as our single free parameter.

For the EFT of GR with dimension-eight operators, Ref.~\cite{Sennett:2019bpc} focused on the
orbital effects (i.e., instead of finite-size effects) and performed
Bayesian model selection using the two lowest-mass BBHs events of the second-observig run of
the LIGO-Virgo Collaboration, notably GW151226 and GW170608. They found that the
data disfavor the appearance of new physics on distance scales around $\ell_{\rm qEFT} \sim 150$ km.

The QNMs of nonrotating BHs where calculated in Refs.~\cite{deRham:2020ejn,Cano:2020cao} in the dimension-six EFT
and in Refs.~\cite{Cardoso:2018ptl,Cano:2020cao} in the dimension-eight EFT.
Reference~\cite{Cano:2020cao} calculated the leading-order BH spin corrections to the QNM spectra.

\section{Methods}
\label{sec:method}

Having reviewed the modified gravity theories that we will consider, we now present
the sequential building blocks for the waveform model we use to test these theories
against GW observations.
We start by reviewing the parametrized ringdown spin-expansion coefficients
(ParSpec) framework~\cite{Maselli:2019mjd} in Sec.~\ref{sec:review_parspec}.
Next, in Sec.~\ref{sec:review_pSEOB}, we review our baseline
parametrized IMR waveform model~\cite{Brito:2018rfr,Ghosh:2021mrv}, and explain how we extend
it to include the ParSpec.
Finally, in Sec.~\ref{sec:theory_specific_qnm}, we show how we can map
theory-specific QNM calculations in modified gravity theories onto the
free coefficients in the ParSpec framework.
Ultimately, this provides us with an IMR waveform model, with the ringdown
portion of the model informed by QNM calculations in specific beyond-GR
theories.

\subsection{The parametrized ringdown spin expansion coefficients framework}
\label{sec:review_parspec}

A general procedure to describe deviations to the QNM frequencies $\omega^{\rm GR}_{\ell m n}$ and
damping times $\tau^{\rm GR}_{\ell m n}$ of BHs of GR is to write,
\begin{subequations}
\label{eq:general_deviation}
\begin{align}
\omega_{\ell m n} &= \omega_{\ell m n}^{\rm GR} \, (1 + \delta \omega_{\ell m n}), \\
\tau_{\ell m n }   &= \tau_{\ell m  n}^{\rm GR}   \, (1 + \delta \tau_{\ell m n }),
\end{align}
\end{subequations}
where $\delta\omega_{\ell m n}$ and $\delta\tau_{\ell m  n}$ are the fractional deformation parameters,
$\ell$ and $m$ are the multipole indices, and $n$ the overtone number.
This type of parametrization\footnote{See also
Refs.~\cite{Glampedakis:2017dvb,Glampedakis:2019dqh,Silva:2019scu} and
Refs.~\cite{Cardoso:2019mqo,McManus:2019ulj,Kimura:2020mrh} for alternative parametrizations.}
was adopted, for instance, in Refs.~\cite{Gossan:2011ha,Meidam:2014jpa,Brito:2018rfr,Carullo:2018sfu,Isi:2021iql,Ghosh:2021mrv}.

The current LVK tests of BH ringdown (see Sec.VII.A
of Ref.~\cite{LIGOScientific:2020tif} or Sec.VIII.A
of Ref.~\cite{LIGOScientific:2021sio}) take a flexible theory-independent
approach towards the inference of $\delta\omega_{\ell m n}$ and
$\delta\tau_{\ell m n}$. These deviations are either assumed to occur
identically across all observed sources or belong to a generic
underlying Gaussian population.
%
%
However, in reality, these parameters depend on the source BH's mass and spin.
Ideally, one would like to explicitly reinstate this dependence, by
(i) introducing deformation parameters which can be determined, once and for
all, from a specific gravity theory (GR included), and
(ii) making it simpler to combine constraints coming from multiple (independently observed) sources.

The ParSpec framework was introduced in Ref.~\cite{Maselli:2019mjd} and can be
used to our purpose. It is an observable-based bivariate expansion of
Eq.~\eqref{eq:general_deviation}, given by
\begin{subequations}
\label{eq:kerr_expansion}
\begin{align}
\omega_{\ell m n} &= \frac{1}{M_{\rm f}} \sum_{j = 0}^{N_{\rm max}} \, \chi_{\rm f}^{j} \omega^{(j)}_{\ell m n} \, \left( 1 + \gamma\, \delta \omega^{(j)}_{\ell m n} \right), \\
\tau_{\ell m n}   &= M_{\rm f}     \sum_{j = 0}^{N_{\rm max}} \, \chi_{\rm f}^{j} \tau^{(j)}_{\ell m n}   \, \left( 1 + \gamma\, \delta \tau^{(j)}_{\ell m n} \right),
\end{align}
\end{subequations}
where $M_{\rm f}$ and $\chi_{\rm f}$ are the detector-frame final mass and spin, respectively;
the quantities $\omega_{\ell m n}^{(j)}$ and $\tau_{\ell m n}^{(j)}$ are dimensionless coefficients of the
expansion in spin for the QNMs of BHs in GR, while $\delta\omega_{\ell m n}^{(j)}$ and $\delta\tau_{\ell m n}^{(j)}$ are
source-independent dimensionless coefficients that characterize the corrections to the GR QNM at
each spin-order,
and $N_{\rm max}$ is the order of the spin expansion.
All source dependence due to a given modified gravity theory is contained in the
dimensionless parameter $\gamma$, which reads
\begin{equation}
\gamma
= \left(\frac{\ell_{\rm th}}{M^{\rm s}_{\rm f}}\right)^{p}
= \left[
\frac{\ell_{\rm th} c^2 (1 + z)}{G M_{\rm f}}
\right]^{p},
\label{eq:def_gamma}
\end{equation}
which depends on the lengthscale parameter $\ell_{\rm th}$ of the specific gravity theory
(non-GR modifications become important at distances $\lesssim \ell_{\rm th}$), and
the exponent $p$ is related to how the non-GR modifications are
added to the Einstein-Hilbert action.
In Eq.~\eqref{eq:def_gamma} we made $\gamma$ dimensionless by the
lengthscale associated with the remnant BH (i.e., its source-frame mass
$M^{\rm s}_{\rm f}$\footnote{Throughout this paper, we maintain the convention of
using the upperscript ``s" to denote source-frame quantities and plain baseline
symbols for detector-frame measurements.}), which we can also write in terms of the detector-frame mass $M_{\rm f}$
through the redshift $z$~\cite{Krolak:1987ofj}. Also, dividing by the factor $G/c^2$ allows us to
express $\ell_{\rm th}$ in physically intuitive metric units.

In principle, a modification to GR would also affect $M_{\rm f}$ and $\chi_{\rm f}$ and the
expansion should be written in terms of the non-GR mass and spin, say $\bar{M}_{\rm f}$ and
$\bar{\chi}_{\rm f}$. If we assume that the non-GR corrections are included perturbatively
(as it is the case with all the theories described in Sec.~\ref{sec:review_theories}),
the modifications to the BH mass and spin can be absorbed into the deviations
parameters $\delta\omega^{(j)}_{\ell m n}$ and $\delta\tau^{(j)}_{\ell m n}$.
This means we can identify the $M_{\rm f}$ and $\chi_{\rm f}$ with their
\emph{corresponding {\rm GR} values}\footnote{For a more detailed discussion, see Appendix A in Ref.~\cite{Maselli:2019mjd}.}.
We will see in Sec.~\ref{sec:pe}  that this assumption is indeed
satisfied in our parameter estimation studies (see, for instance, Fig.~\ref{fig:corner_plot}).

Finally, we remark that in the GR limit ($\gamma \to 0$)
the series~\eqref{eq:kerr_expansion} truncated at $N_{\rm max} = 4$, reproduces with 1\% accuracy
the GR QNMs for BH's spins $\chi_{\rm f} \lesssim 0.7$.
The values of the fitting coefficients $\omega_{\ell m n}^{(j)}$ and $\tau_{\ell m n}^{(j)}$
can be found in Ref.~\cite{Maselli:2019mjd}.
In Ref.~\cite{Carullo:2021dui}, the fitting coefficients were calculated up to $N_{\rm max} = 9$,
which extend the validity of the spin-expansion up to $\chi_{\rm f} \lesssim 0.99$.
As we will discuss in Sec.~\ref{sec:pe}, the expansion to $N_{\rm max} = 4$ is
sufficient for our purposes. For convenience we list the coefficients in the case of
GR in Table~\ref{tab:ref_theories_qnms}.

\subsection{The parametrized waveform model}
\label{sec:review_pSEOB}

We now describe the waveform model used in our paper to infer properties of a
BBH ringdown.
As in Refs.~\cite{Brito:2018rfr,Ghosh:2021mrv}, we use an IMR BBH waveform
model where the complex-valued frequencies describing the remnant object
are left additionally free and estimated directly from the data.

The GW signal from a quasicircular BBH can be described  in GR by a unique set of
parameters $\bm{\theta}$, that includes the masses and spins of the two BHs,
$(m_1,\, m_2,\, \mathbf{S}_1,\, \mathbf{S}_2)$, the sky location determined by the
luminosity distance $D_L$,
right ascension $\alpha$ and declination $\delta$,
and the orientation of the binary given by the inclination $\iota$ and polarization $\psi$
angles.
The set is completed by the choice of a reference time $t_0$ and phase
$\phi_0$. If we further assume that the spins of the individual BHs are
restricted to be aligned or anti-aligned (for short, aligned) to the unit vector
perpendicular to the orbital plane ($\mathbf{\hat{L}}$), we reduce the six
components of the spins to just two, $\chi_i \equiv \mathbf{S}_i\cdot \mathbf{\hat{L}} /m_i^2$ with $i = 1,2$,
and our entire parameter set from 15 to 11.
Let us also define some additional parameters and set some conventions that
will be useful in our analysis later, namely, the total mass
$M=m_1+m_2$,
the chirp mass
$\mathcal {M}=(m_{1}m_{2})^{3/5}/(m_{1}+m_{2})^{1/5}$,
the asymmetric mass ratio $q=m_1/m_2$, with the convention $m_1 \geqslant m_2$ (and thus $q \geqslant 1$),
and the symmetric mass ratio of the binary, $\nu = m_1m_2/(m_1+m_2)^2$. Note that for BHs $-1 \leqslant \chi_i \leqslant 1$.

For the polarizations of the GW signal (in the observer's frame) we have
\begin{equation}
h_+(\iota,\varphi_0;t ) - i h_\times(\iota,\varphi_0;t) = \sum_{l, m} {}_{-\!2}Y_{l m}(\iota,\varphi_0)\, h_{l m}(t),
\end{equation}
where ${}_{-\!2}Y_{l m}(\iota,\varphi_0)$ are the $-2$ spin-weighted spherical
harmonics.
As our baseline model, that is, the GR model upon which non-GR modifications
are added, we use the computationally efficient (time-domain) multipolar
waveform model for quasicircular spin-aligned BBHs described in
Ref.~\cite{Mihaylov:2021bpf}, which contains the modes, $(l, |m|)=(2,2),(2,1)$, $(3,3)$, $(4,4)$,
and $(5,5)$. Such a model was built by applying the post-adiabatic approximation~\cite{Nagar:2018gnk}
to the multipolar spin-aligned EOB waveform model of Refs.~\cite{Bohe:2016gbl,Cotesta:2018fcv} (henceforth we refer to our baseline model as \SEOB{}\footnote{This waveform
model is available in \texttt{LALSuite}~\cite{lalsuite} as the \texttt{SEOBNRv4HM\_PA}
waveform approximant.}).

An accurate description of the merger is incorporated through calibration with
NR simulations, as described in Refs.~\cite{Bohe:2016gbl,Cotesta:2018fcv}, along with
information for the merger and ringdown phases, from BH perturbation theory.
The merger-ringdown waveform, $h_{l m}^\textrm{merger-RD}$, is then stitched to
inspiral-plunge waveform, $h_{l m}^\textrm{insp-plunge}$ at a certain time $t = t^{\textrm{match}}_{l m}$, as
\begin{equation}
h_{l m}(t) =
h_{l m}^\textrm{insp-plunge}\, \Theta(t_\textrm{match}^{l m} - t)
+ h_{l m}^\textrm{merger-RD}\,\Theta(t-t_\textrm{match}^{l m})\,,
\end{equation}
where $\Theta(t)$ is the Heaviside step function.
The merger-ringdown waveform is expressed as an exponentially damped
sinusoid~\cite{Bohe:2016gbl,Cotesta:2018fcv}
\begin{equation}
\label{RD}
h_{l m}^{\textrm{merger-RD}}(t) = \nu \ \tilde{A}_{l m}(t)\ e^{i \tilde{\phi}_{l m}(t)} \ e^{-i \sigma_{l m 0}(t-t^{\textrm{match}}_{l m})}\,,
\end{equation}
where
\begin{align}
\sigma_{l m 0} \equiv {\rm Re}(\sigma_{l m 0}) + i \, {\rm Im}(\sigma_{l m 0})
= \omega_{l m 0} - \frac{i}{\tau_{l m 0}}\,,
\end{align}
are the complex frequencies of the fundamental ($0$-th overtone) QNMs of the remnant BH.
The functions $\tilde{A}_{l m}(t)$ and $\tilde{\phi}_{l m}(t)$ are defined
in Ref.~\cite{Bohe:2016gbl,Cotesta:2018fcv}.

In the $\SEOB$ model~\cite{Bohe:2016gbl,Cotesta:2018fcv}, the complex frequencies
$\sigma_{\ell m 0}$ are computed by first determining the final mass and spin from
estimates of the initial masses and spins through NR-fitting-formulas~\cite{Taracchini:2013rva,Hofmann:2016yih},
and then converting them to the complex frequencies using BH perturbation
theory-inspired analytical fits~\cite{Berti:2005ys,Berti:2009kk}.
Hence,
\begin{subequations}
\begin{align}
\omega_{\ell m 0}^{\rm GR} &\equiv \omega^{\rm GR}_{\ell m 0}(m_1, m_2, \chi_1, \chi_2)\,,
\\
\tau _{\ell m 0}^{\rm GR} & \equiv \tau^{\rm GR}_{\ell m 0}(m_1, m_2, \chi_1, \chi_2)\,,
\end{align}
\end{subequations}
where $(\omega^{\rm GR}_{\ell m 0}, \tau^{\rm GR}_{\ell m 0})$ refer to the GR QNM predictions in the baseline \SEOB{} model.
In this paper, we replace these GR predictions with QNM
frequencies defined through the ParSpec framework introduced in
Sec.~\ref{sec:review_parspec} (see Eqs.~\eqref{eq:kerr_expansion}). Hence,
\begin{subequations}
\begin{align}
\omega_{\ell m 0} & \equiv  \omega_{\ell m 0}(m_1, m_2, \chi_1, \chi_2,\ell_{\rm th}, \{\delta \omega_{\ell m 0}^{(j)}\}),\\
\tau_{\ell m 0}   & \equiv \tau_{\ell m 0} (m_1, m_2, \chi_1, \chi_2, \ell_{\rm th}, \{\delta \tau_{\ell m 0}^{(j)}\}),
\end{align}
\end{subequations}
where the mass and spin of the remnant object $(M_{\rm f},\chi_{\rm f})$ are themselves functions of
$(m_1,\, m_2,\, \chi_1,\, \chi_2)$~\cite{Taracchini:2013rva,Hofmann:2016yih}, and we fix $p$
to a certain theory-specific value. Additionally, the frequencies depend on the ParSpec coefficients $\{\delta
\omega_{\ell m 0}^{(j)},\delta \tau_{\ell m 0}^{(j)}\}$, and the lengthscale $\ell_{\rm th}$.

Using this parametrized waveform model, which we call $\pSEOB$,
we infer bounds on our non-GR parameter $\ell_{\rm th}$ for the specific cases of
modified gravity theories presented Sec.~\ref{sec:review_theories}.
We detail our results in Sec.~\ref{sec:results}.


\subsection{From theory-independent to theory-specific QNM results}
\label{sec:theory_specific_qnm}

Let us now establish the connection between the theory-independent framework of the
$\pSEOB$ waveform model and the theory-specific QNM calculations. In this paper we restrict ourselves
to the leading and next-to-leading order terms in the ParSpec expansion, as well as to the fundamental QNM
$(\ell, m, n) = (2,\, 2,\, 0)$. For this reason, for simplicity, we omit the subscripts hereafter
and rewrite $\omega_{\ell m n}$ and $\tau_{\ell m n}$, given by Eqs.~\eqref{eq:kerr_expansion} as,
\begin{subequations}
\label{eq:qnm_nongr_iso}
\begin{align}
    M_{\rm f}\, \omega &= \gamma \left [ \delta\omega^{(0)} \omega^{(0)} + \chi_{\rm f}\, \delta\omega^{(1)} \omega^{(1)} \right ]
    + \sum_{j=0}^{N_{\rm max}} \chi^{j}_{\rm f}\, \omega^{(j)} \,,
\\
    \frac{\tau}{M_{\rm f}}   &= \gamma \left [ \delta\tau^{(0)} \tau^{(0)} + \chi_{\rm f}\, \delta\tau^{(1)} \tau^{(1)} \right ]
    + \sum_{j=0}^{N_{\rm max}} \chi^{n}_{\rm f}\, \tau^{(j)}\,,
\end{align}
\end{subequations}
where we pull out from the sum all non-GR corrections, restricting
ourselves to the nonspinning ($j=0$) and linear-order in spin ($j=1$)
corrections to the GR QNMs.
The QNMs associated to the higher ($\ell, |m|$)-modes listed in Sec.~\ref{sec:review_pSEOB} are kept with their GR values.

\begin{table*}[t]
\begin{tabular}{c | c }
\hline
\hline
                        & GR~\cite{Maselli:2019mjd} \\
\hline
$\omega^{(0)}$    & 0.3737                    \\
$\tau^{(0)}$      & 11.2407                   \\
$\omega^{(1)}$    & 0.1258                    \\
$\tau^{(1)}$      & 0.2522                    \\
\hline
\hline
\end{tabular}
\quad
\begin{tabular}{c | c  c  c  c}
\hline
\hline
                              & EdGB ($p=4$)~\cite{Blazquez-Salcedo:2016enn,Pierini:2021jxd} & dCS ($p=4$)~\cite{Wagle:2021tam} & cubic EFT ($p=4$)~\cite{Cano:2021myl} & quartic EFT ($p=6$)~\cite{Cano:2021myl} \\
\hline
$\delta\omega^{(0)}$    & 0.0107                                               & 3.1964                   & $-$0.5813                     & $-$0.2114                       \\
$\delta\tau^{(0)}$      & 0.0044                                               & 6.3619                   & $-$0.2114                     & $-$0.6070                       \\
$\delta\omega^{(1)}$    & $-$0.2480                                            & 41.199                   & 6.4439                        & $-$1.5263                       \\
$\delta\tau^{(1)}$      & $-$1.1014                                            & 794.66                   & 265.12                        & 171.35                          \\
\hline
\hline
\end{tabular}
\caption{Summary of theory-specific QNM calculations.
We summarize each theory we consider together with: the exponent $p$ at
which their QNM-modification enters, the corresponding modifications to the
oscillation frequency $\delta \omega^{(n)}_{220}$ and decay time $\delta \tau^{(n)}_{220}$, and the
references from which we used the results from.
We also include for comparison the GR coefficients, up to $j=1$, obtained in Ref.~\cite{Maselli:2019mjd}.
The remaining GR coefficients for $1< j \leqslant 4$, for which their non-GR counterpart cannot be determined as of yet for
the theories under consideration, can be found in Table~I of Ref.~\cite{Maselli:2019mjd}.
}
\label{tab:ref_theories_qnms}
\end{table*}

How can we determine the beyond-GR corrections? In GR, comparison between the numerically determined
Kerr QNMs against the fitting formula~\eqref{eq:qnm_nongr_iso} fixes the GR expansion coefficients
$\omega^{(j)}$ and $\tau^{(j)}$.
We can proceed in a similar way with QNMs calculated in the context of a non-GR theory.
In particular, in the literature, we can already find fitting formulas relating
the QNMs to the BHs mass, spin and lengthscale $\ell_{\rm th}$, the latter being
specific to each theory, up to $j=1$ in the spin expansion (see Table~\ref{tab:ref_theories_qnms}).
The idea is then to compare these formulas against Eq.~\eqref{eq:qnm_nongr_iso}
to fix $p$, $\delta\omega^{(j)}$, and $\delta\tau^{(j)}$.
Because QNMs of rotating BHs in modified gravity theories are not known to
all spin values, we can expect that the $j=1$ coefficients to change as
calculations beyond-leading order in spin are accomplished in the future.
That is not the case for the $j=0$ coefficients and the situation is the same as in GR,
in which the $j=0$ coefficients are simply the QNMs of the Schwarzschild BH.

In the end, the $\pSEOB$ waveform model with theory-specific
QNMs has only $\ell_{\rm th}$ as a free beyond-GR parameter.
We emphasize that our procedure is different from that of
Ref.~\cite{Carullo:2021dui} which, for a given value of $p$, varied all $\ell_{\rm th}$,
$\delta\omega^{(j)}$, and $\delta\tau^{(j)}$ parameters, and then proceeded
to use the posteriors on $\ell_{\rm th}$, considering up to $j=2$ in
the GR deformation coefficients, and remaining agnostic about the underlying
theory which would predict the modifications to the QNMs.
We will see in Sec.~\ref{sec:results} that adding theory-specific
information to the ParSpec coefficients \emph{can lead to different interpretations
of the bounds on $\ell_{\rm th}$, even for different theories that predict the same
value of the exponent $p$.}

As we have seen in Sec.~\ref{sec:review_theories}, QNMs of slowly rotating
BHs in modified gravity theories can belong to two families depending on how
they behave under a parity transformation: axial and polar. Which one do
we use to match with Eqs.~\eqref{eq:qnm_nongr_iso}?
To answer this question one has to work with a chosen theory and perform a
translation between the metric perturbations $h_{\mu\nu}$ in the
Regge-Wheeler-Zerilli gauge~\cite{Regge:1957td,Zerilli:1970se} and connect it with the transverse-traceless gauge
used to described GWs (see, for instance,~Ref.~\cite{Maggiore:2018sht}, Chapter~12),
In GR, both axial and polar QNMs are isospectral and hence which QNM we use to
model the ringdown makes no difference.
In beyond-GR theories, isospectrality is in general broken (see
in Ref.~\cite{Hui:2021cpm} for a counterexample).
Thus, how axial and polar gravitational QNMs appear in the GW signal has to be
answered on a theory-by-theory basis.
This is outside the scope of this paper and here we take the more pragmatic
approach of simply choosing the \emph{least damped gravitational mode between
the two parities.}
%
Underlying this choice, are the assumptions that either (i) the
least-damped QNM is also the one excited with largest amplitude or (ii) that
QNMs of both parities are excited with comparable amplitudes, and one
of the modes decays sufficiently fast to not appear in the ringdown.
We performed the mapping between theory-specific QNM calculations and the ParSpec framework
under the hypothesis above, for the theories listed in Sec.~\ref{sec:review_theories}.
We summarize our results in Table~\ref{tab:ref_theories_qnms} and
leave the details of our calculations to Appendix~\ref{app:map_details}.

\begin{figure}[htb]
\includegraphics[width=\columnwidth]{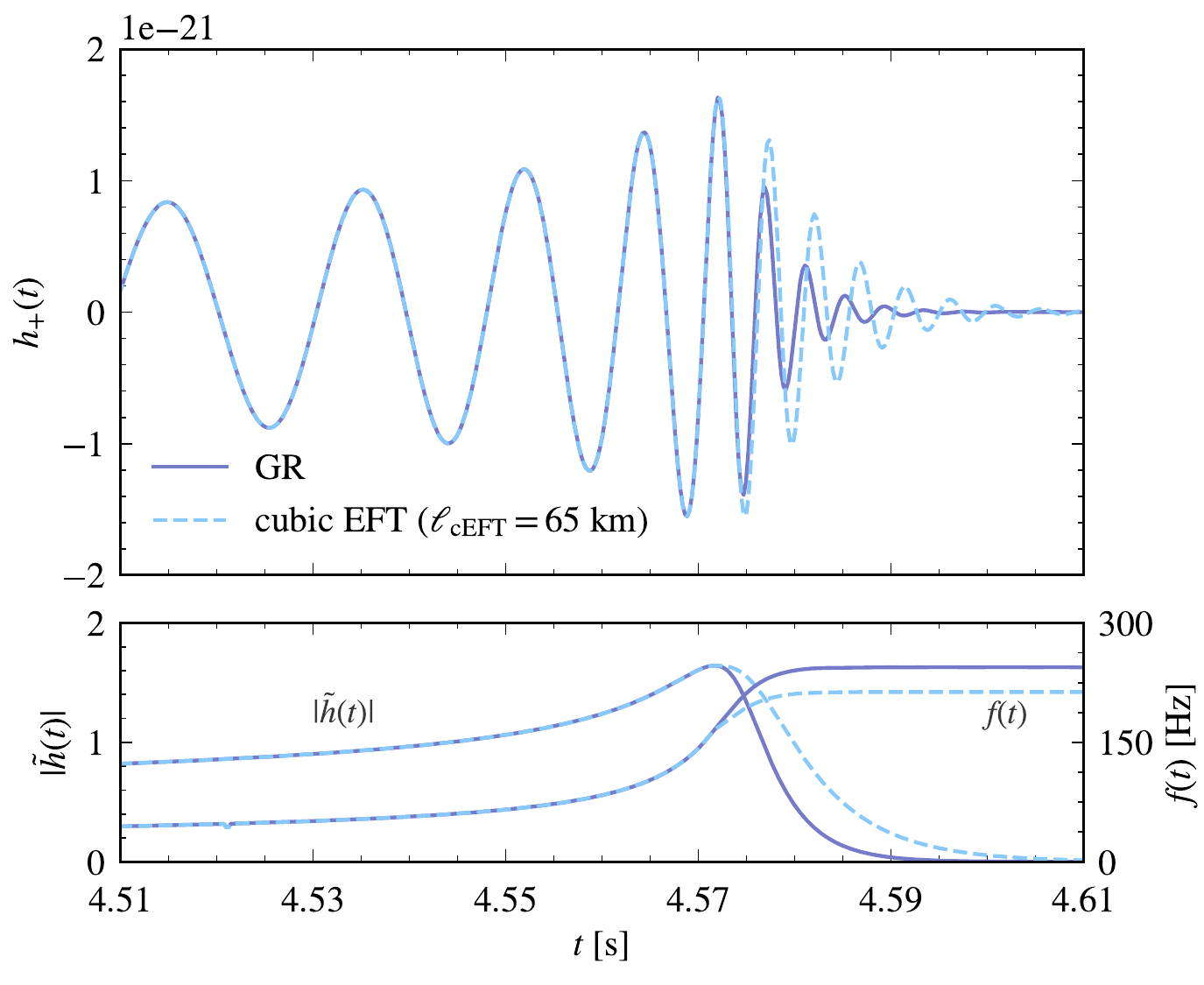}
\caption{Gravitational-wave signal with GW150914-like parameters
for both GR (solid line) and cubic EFT of GR (dashed line) with the leading-order
$n=0$ modifications to the fundamental QNM, for $\ell_{\rm cEFT} = 65$~km.
The former is computed with the \SEOB{} model, while the latter with the
\pSEOB{} model, with ringdown modifications according to the results in
Table~\ref{tab:ref_theories_qnms}.
Top panel: the $+$ polarization $h_{+}(t)$. Bottom panel: the GW amplitude
$|\tilde{h}(t)|$ (left axes) and the instantaneous frequency $f(t)$ (right axes).
}
\label{fig:example_waveform}
\end{figure}

In Fig.~\ref{fig:example_waveform} we show an illustrative waveform for GR (solid line; using the \texttt{SEOBNR} model) and in the cubic EFT of gravity (dashed line; using the \pSEOB{} model),
including the leading-order $j = 0$ deformations to the fundamental QNM (see Table~\ref{tab:ref_theories_qnms}).
We choose binary parameters similar to GW150914,  but nonspinning, with detector-frame masses
$m_1 = 39\msun$ and $m_2 = 31\msun$.
The top panel shows the $+$ GW polarization $h_{+}$ in both theories,
while the bottom panel shows the amplitude $|h| = (h_{+}^2 + h_{\times}^2)^{1/2}$
and instantaneous frequency $f$, all as functions of time $t$.
The signals are identical up to the merger, specifically, the time $t_\textrm{match}$
defined in Sec.~\ref{sec:review_pSEOB}, after which they differ during the ringdown.
By construction, the ringdown lasts longer for the cubic EFT of GR waveform and
with a smaller instantaneous frequency (see the bottom panel) due to the negative value of the
$\delta\omega^{(0)}$ coefficient in this theory.

\section{Parameter inference and validity of our bounds}
\label{sec:pe}

In this section, we first provide a basic outline of the Bayesian formalism that we use to
infer the properties of the underlying GW signal; then, we identify the most promising
events from the catalog of LVK GW observations to base our analyses on. Finally,
we discuss how we can interpret our results after taking into account the region of validity of
the non-GR theories that we are considering, which are EFTs.

\subsection{Bayesian formalism}
If we assume that a GW signal observed in detector data $d$ is accurately
described by our waveform model \pSEOB{}, we can infer the parameters of the
model, $\bm{\lambda}$, given the hypothesis $\mathcal{H}$, using Bayes' theorem,
\begin{equation}
P(\bm{\lambda} \vert d, \mathcal{H}) =
\frac{p(\bm{\lambda} \vert \mathcal{H}) \, \mathcal{L}(d \vert \bm{\lambda},\mathcal{H})}{E(d \vert \mathcal{H})}\,,
\label{eq:bayes}
\end{equation}
where $P(\bm{\lambda} \vert d, \mathcal{H})$ is the posterior probability distribution,
$p(\bm{\lambda} \vert \mathcal{H})$ the prior,
$\mathcal{L}(d \vert \bm{\lambda},\mathcal{H})$ the likelihood, and
$E(d \vert \mathcal{H})$ the evidence.
The set of parameters, $\bm{\lambda}$ is a union of the GR waveform
model parameters~$\bm{\theta}$ (see Sec.~\ref{sec:review_pSEOB})
and $\ell_{\rm th}$, the only non-GR parameter in this problem which, we recall, sets
the characteristic lengthscale in which deviations from
GR become relevant in each of the theories described in
Sec.~\ref{sec:review_theories}.

Assuming stationary Gaussian noise, we can write the (log) likelihood function as,
\begin{equation}
\ln \mathcal{L}(d \vert \bm{\lambda},\mathcal{H}) \propto
- \tfrac{1}{2}
\langle
d - h(\bm{\lambda}) \vert d - h(\bm{\lambda})
\rangle\,,
\end{equation}
with the noise-weighted inner product $\langle \cdot | \cdot \rangle$ defined as,
\begin{equation}
\langle A | B \rangle =
2
\int_{f_{\rm low}}^{f_{\rm high}} \, \dd f \,
\frac{\tilde{A}^{\ast}(f) \tilde{B}(f) + \tilde{A}(f) \tilde{B}^{\ast}(f)}{S_{n}(f)}\,,
\end{equation}
where $\tilde{A}(f)$ is the Fourier transform of $A(t)$, the asterisk denotes
complex conjugation, $S_{n}(f)$ is the power spectrum density of the
detector, and $[f_{\rm low}, f_{\rm high}]$ span the detector sensitivity frequency band.
Assuming a specific prior distribution for our parameters (discussed further in the next section), we stochastically
sample over the parameter space using a Markov-Chain Monte Carlo algorithm as implemented in
\texttt{LALInferenceMCMC}~\cite{Rover:2006ni,vanderSluys:2008qx}, as part of the \texttt{LALInference} software suite~\cite{Veitch:2014wba,lalsuite}.
We subsequently marginalize over the remaining parameters to obtain the
posterior probability distribution function~(PDF) on $\ell_{\rm th}$ i.e., $P(\ell_{\rm th} \vert d,\mathcal{H})$,
our main parameter of interest.

For $N$ independent GW observations $\{d_j\}$, $j=1,...,N$, each characterized
by a PDF $P_j(\ell_{\rm th} \vert d_j,\mathcal{H})$, the
joint posterior can be written as:
\begin{equation}
P(\ell_{\rm th} | \{d_j\},\mathcal{H}) = p(\ell_{\rm th}) \prod_{j=1}^{N} \frac{P_j(\ell_{\rm th} | d_j,\mathcal{H})}{p_j(\ell_{\rm th}|\mathcal{H})}\,.
\label{eq:cumulative_dist_ell}
\end{equation}
where $p_j(\ell_{\rm th} | \mathcal{H})$ are the priors used for each observation, $p(\ell_{\rm th})$ is an overall prior, and we assume that the value of $\ell_{\rm th}$ is shared among all events.
Since we assume a uniform prior on $\ell_{\rm th}$, the joint posterior is
equal to the joint likelihood. Hereafter, we will drop the explicit usage of
$\mathcal{H}$.

\subsection{Priors}

The prior distribution functions on the GR parameters are assumed to be
uniform over the component masses, $(m_1, m_2)$, isotropically distributed on a
sphere in the sky for the source location with $p(D_L) \propto D_L^2$, and isotropic
on the binary orientation, $p(\iota, \psi, \phi_0) \propto \sin\iota$. For the spins
$(\chi_1, \chi_2)$, we assume a prior uniform and isotropic in the spin magnitudes\footnote{This spin-prior choice can be specified in \texttt{LALInference} using the option \texttt{alignedspin-zprior.}}.

Among our non-GR parameters $\{\ell_{\rm th}, \delta \omega^{(j)}\delta \tau^{(j)}\}$,
as already mentioned in the previous section, we hold $\{\delta \omega^{(j)},\delta \tau^{(j)}\}$
fixed to theory-specific predictions, and only allow $\ell_{\rm th}$ to vary freely.
We assume a uniform prior on $\ell_{\rm th}$,
which ranges between $\ell_{\rm th} = 0$~km and
$\ell_{\rm th} \sim 100 - 300$~km, the specific value chosen to ensure that the
marginalized posterior distributions on this parameter do not rail against the
prior's maximum value.
The lower limit is set by the fact that the modified gravity theories we consider
all have $p$ even and hence we can assume $\ell_{\rm th} > 0$ without loss of
generality.

\subsection{Events selection}

The \pSEOB{} model, as described in Sec.~\ref{sec:review_pSEOB}, is an
IMR model that infers the properties of the underlying GW signal,
including (independently) its ringdown properties, using the Bayesian formalism
above. Naturally, the most promising candidates for our analyses are high-mass
\emph{and} loud GW observations with a significant signal-to-noise ratio (SNR)
in the inspiral and post-merger stages to break the degeneracy between the total mass
and the QNM frequencies.
The latest LVK GW catalog~\cite{LIGOScientific:2021djp} reported 90 observed
signals not all of which are relevant for our BH ringdown analysis. In fact, in
the accompanying paper~\cite{LIGOScientific:2021sio} on tests of GR, the \texttt{pSEOBNRv4HM}
~\cite{Brito:2018rfr,Ghosh:2021mrv} analysis~\footnote{See, in particular, Sec.VIII A.2
in Ref.~\cite{LIGOScientific:2021sio}}, which is most similar to the \pSEOB{} model
presented in this paper, identified two events which provided the strongest
bounds on the measurements of the dominant $(220)$ QNM:
GW150914~\cite{LIGOScientific:2016aoc} and
GW200129~\cite{LIGOScientific:2021djp}.
These two events, with a total (source-frame) mass of $65 \Mo$ and $63.4 \Mo$ respectively, are
extremely similar in their source properties. These are also two of the loudest
BBH signals observed to date with a total network SNR of 24 and 26.8,
respectively.
Moreover, and what is more relevant for our analysis, are their post-inspiral
(merger-ringdown) SNRs which are both $\approx 16$ (see the columns for
$\rho_{\text{post-insp}}$ in Table III of Ref.~\cite{LIGOScientific:2019fpa} and
Table IV of Ref.~\cite{LIGOScientific:2021sio}).
In this paper, we are going to focus on these two GW events as our probes of
the BH ringdown in modified theories of gravity.

The parameter inference in this paper follows configurations identical to the
ones used on these events for the \texttt{pSEOBNRv4HM} analysis in
Ref.~\cite{LIGOScientific:2021sio}. GW150914 was a 2-detector (Hanford-Livingston) event while
GW200129 was 3-detector (Hanford-Livingston-Virgo).
We consequently use the same strain data $h(t)$, detector
power-spectral-densities $S_n(f)$ and calibration envelopes as were used for
the analyses in Ref.~\cite{LIGOScientific:2021sio}.

In Sec.~\ref{sec:results}, we enumerate through the different theories and
outline the main results. Whenever possible, we also combine results from
both events to obtain the strongest possible bound on $\ell_{\rm th}$.

\subsection{EFT interpretation of our results}
\label{sec:remarks}

There are two conditions that we must verify before we can confidently claim to
have placed a constraint on $\ell_{\rm th}$. First, as we have explained in Sec.~\ref{sec:review_theories}, all theories that we
consider must be interpreted as an EFT, meaning that they should be
considered valid only below an energy scale, or equivalently, above a lengthscale.
As a cutoff lengthscale for the validity of the EFT we use,
\begin{equation}
\Lambda_{\rm EFT} (\varepsilon, \mathfrak{m}) = \varepsilon \, \frac{G \mathfrak{m}}{c^2} \,,
\label{eq:def_cutoff}
\end{equation}
where $\varepsilon$ is a dimensionless number and $\mathfrak{m}$ is the median
value of one of the mass scales involved in the problem.
We note that $\Lambda_{\rm EFT}$ has dimensions of length and hence can be compared to each theory's fundamental lengthscale $\ell_{\rm th}$.
Here we explore the range $\varepsilon \in [0, 1]$, but following
Refs.~\cite{Nair:2019iur,Perkins:2021mhb,Lyu:2022gdr} we quote our final
results using $\varepsilon = 1/2$, but we stress that there is no
fundamental justification for this choice.

Under these assumptions, we will say that a bound has been placed on $\ell_{\rm th}$, if most of the PDF $P(\ell_{\rm th} | d)$ support is in the interval
$[0, \, \Lambda_{\rm EFT}(1/2, \mathfrak{m})]$.
In practice, this can be quantified through the cumulative distribution function
(CDF) associated with the marginalized posterior distribution $P(\ell_{\rm th} | d)$, namely
\begin{equation}
P(\ell_{\rm th} \leqslant \ell^{\rm max}_{\rm th} | d) = \int_{0}^{\,\ell^{\rm max}_{\rm th}} \dd \ell'\, P(\ell' | d).
\label{eq:def_cdf}
\end{equation}
For instance, we require that for a bound at 90\% credible level to be placed on $\ell_{\rm th}$ that
\begin{equation}
P(\ell_{\rm th} \leqslant \Lambda_{\rm EFT} | d) \geqslant 0.9,
\quad \textrm{(EFT bound)},
\label{eq:eft_bound}
\end{equation}
where we let $\ell^{\rm max}_{\rm th} = \Lambda_{\rm EFT}$ in Eq.~\eqref{eq:def_cdf}, and
likewise for other credibility percentiles.

Second, as already emphasized in Ref.~\cite{Maselli:2019mjd}, the ParSpec
formalism is by construction perturbative. This means that the non-GR
deformation parameters are small, that is,
\begin{equation}
\gamma \, \delta \omega^{(j)} \ll 1,
\,\, \textrm{and} \,\,
\gamma \, \delta \tau^{(j)} \ll 1, \quad \textrm{(ParSpec bound)},
\label{eq:parspec_bound}
\end{equation}
for all orders $j$ in the expansion in dimensionless spin $\chi_{\rm f}$ and where $\gamma$
was defined in Eq.~\eqref{eq:def_gamma}.
We also construct posterior distributions for these parameters and check if
most of their support is concentrated to a domain with values much smaller than
unity.

Another question we must consider is the following: what is the mass $\mathfrak{m}$ that we should use
in Eq.~\eqref{eq:def_cutoff}?
In Refs.~\cite{Nair:2019iur,Perkins:2021mhb,Lyu:2022gdr}, which attempted to
constrain dCS and EdGB theories with the \emph{inspiral} part of
the GW signal alone, it was natural to choose the secondary's mass $m_2$ as the most
conservative choice, since it is by definition the smaller component mass and hence
places the lowest cutoff scale $\Lambda_{\rm EFT}$ for the validity of either of these theories as an EFT.

In our problem, the answer is not as clear. On the one hand, since we are
interested in the ringdown part of the signal, it is natural to use the
final mass $M_{\rm f}$ to compute $\Lambda_{\rm EFT}$.
On the other hand, one may argue that the modified gravity theory under
consideration should be able to predict a full inspiral, merger, and ringdown
of the BBH before we can even make such a test, and thus the
same, more conservative choice $\gm = m_2$ should be used.
Here we adopt a pragmatic approach to this issue and consider \emph{both}
masses, $m_2$ and $M_{\rm f}$, to determine $\Lambda_{\rm EFT}$.
Specifically, we will use the median value of the marginalized PDF of these
masses.
We then compare how different assumptions yield to different
interpretations of the results of our parameter estimation.

\section{Results using LIGO-Virgo events}
\label{sec:results}

\subsection{Einstein-dilaton-Gauss-Bonnet gravity}
\label{sec:results_edgb}

\begin{figure}[t]
\includegraphics[width=\columnwidth]{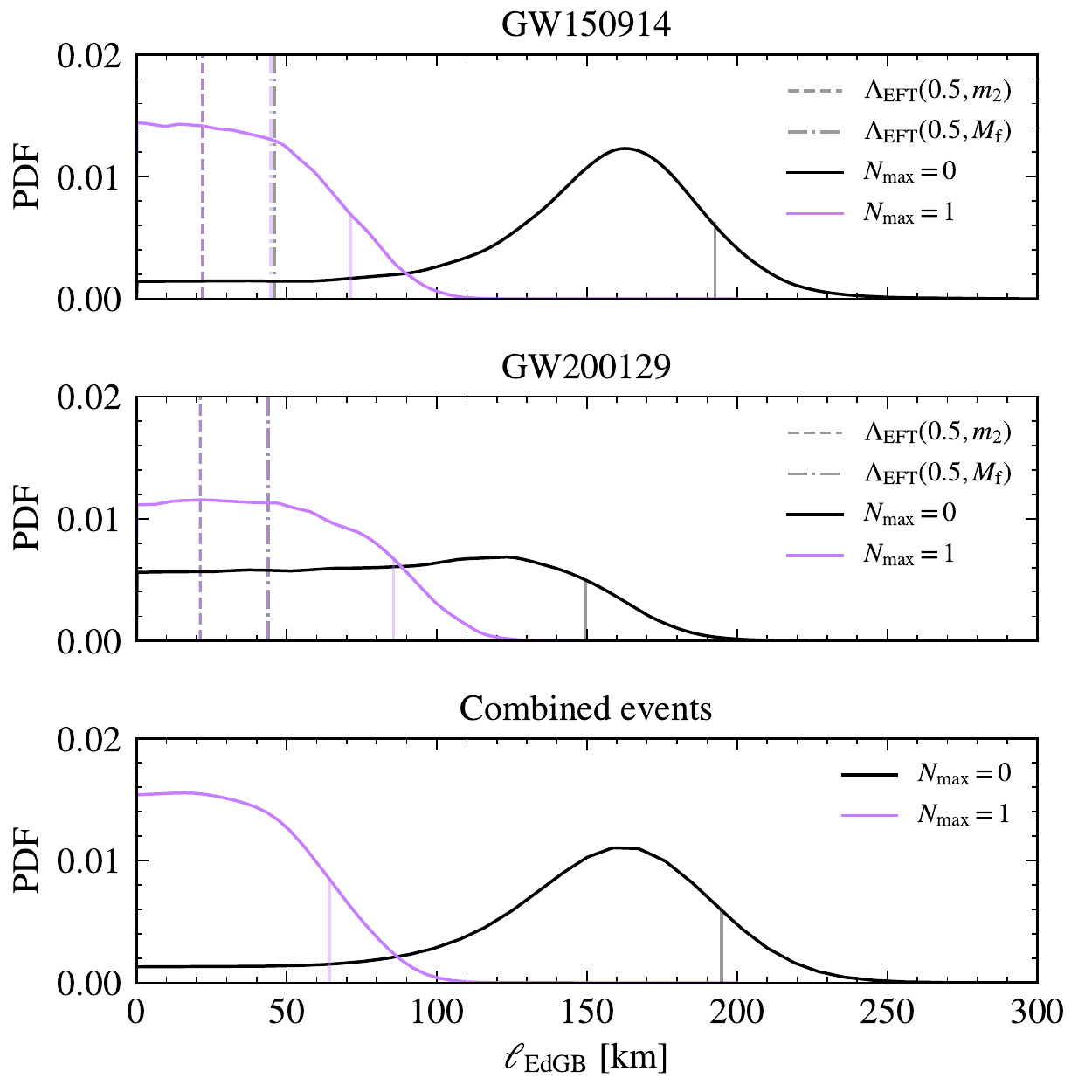}
\caption{Posterior distribution function on coupling constant $\ell_{\rm EdGB}$ in EdGB gravity
for GW150914 (top panel), GW200129 (middle panel)
and from combining results (bottom panel).
In all panels, different line colors correspond to the inclusion ($N_{\rm max} = 1$) or not ($N_{\rm max} = 0$)
of the linear-in-spin QNM correction.
The joint posteriors are shown for illustrative purposes only. As we explain in Fig.~\ref{fig:edgb_cdf} and
in the main text, our analysis of these events fails to satisfy the EFT bound~\eqref{eq:eft_bound}. We mark
with solid vertical lines the $90\%$ upper credible intervals, while the dashed and dot-dashed lines correspond to the EFT bound, $\Lambda_{\rm EFT}(0.5, m_2)$ and
$\Lambda_{\rm EFT}(0.5, M_{\rm f})$, respectively. We see that the $90\%$ upper credible level lines
are located to right of the EFT cutoffs. This means that the EFT bound is not satisfied.
}
\label{fig:edgb_exec_sum}
\end{figure}

\begin{figure}[h]
\includegraphics[width=\columnwidth]{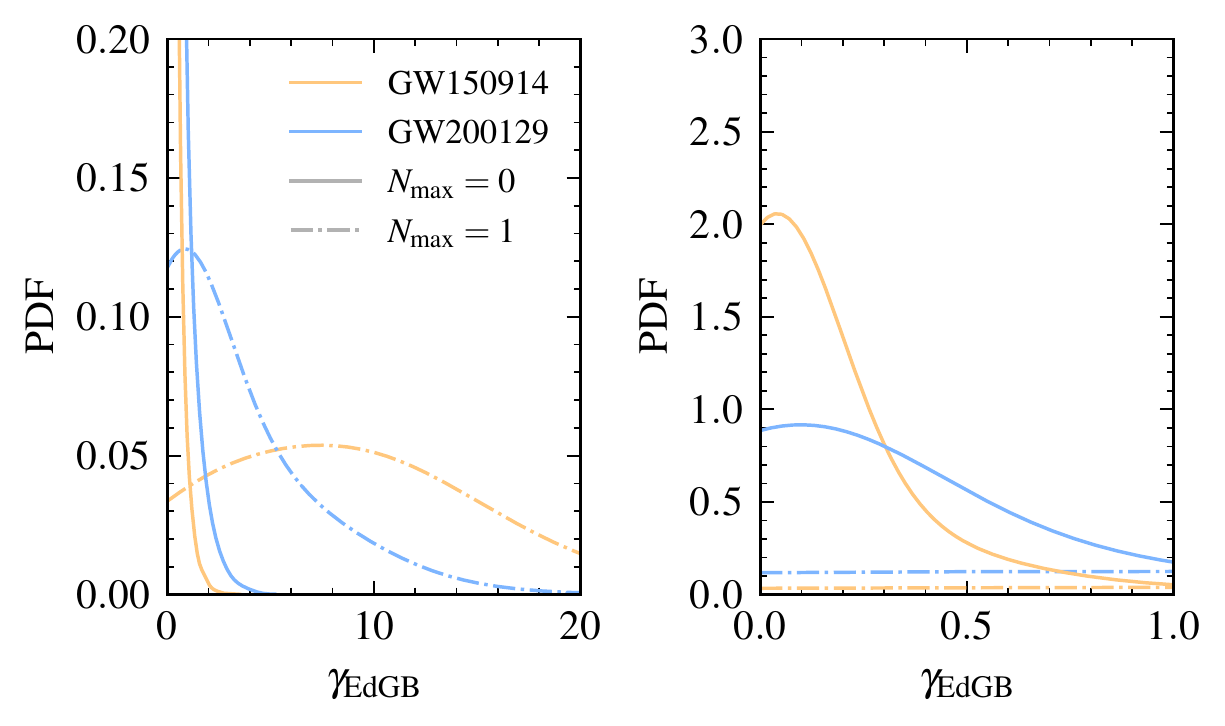}
\caption{Posterior distribution function on the dimensionless parameter $\gamma_{\rm EdGB}$,
defined in Eq.~\eqref{eq:def_gamma_gb}. Different line colors distinguish between events, while
different line styles distinguish between different $N_{\rm max}$.
We see that this parameter which controls the ParSpec expansion in EdGB gravity
does have maximum support at $\gamma_{\rm EdGB} = 0$. This shows that our model is consistent with GR.
The right panel shows the PDFs in the range $0 \leqslant \gamma_{\rm EdGB} \leqslant 1$. Note how
the curves are flat in this range for both events and when $N_{\rm max} = 1$.}
\label{fig:gamma_gb}
\end{figure}

We start with EdGB gravity.
In Fig.~\ref{fig:edgb_exec_sum} we show the marginalized PDFs of the
coupling constant $\ell_{\rm GB}$, for GW150914 (top panel) and GW200129
(middle panel), with and without the spin corrections to the $(2,2,0)$ QNM.
The bottom panel shows the joint posterior obtained by combining both events.
We see, for both events, the $N_{\rm max} = 0$ posteriors are
characterized by a peak away from zero.
This does not mean that we are inferring a deviation from GR.
We recall that the deviations from GR in the ParSpec framework are controlled
by the dimensionless parameter $\gamma$, which here reads,
\begin{equation}
    \gamma_{\rm EdGB} = \left(\frac{c^2 \ell_{\rm EdGB}}{ G M^{\rm s}_{\rm f}}\right)^4\,.
    \label{eq:def_gamma_gb}
\end{equation}
As shown in Fig.~\ref{fig:gamma_gb}, $\gamma_{\rm EdGB}$ does indeed have a posterior
distribution with largest support at zero, indicating consistency between
the underlying signal and GR.
We also observe that the inclusion of the spin corrections (i.e., the curves with $N_{\rm max} = 1$)
displaces the posteriors distributions towards smaller values of $\ell_{\rm EdGB}$ (see Fig.~\ref{fig:edgb_exec_sum}),
and larger values of $\gamma_{\rm EdGB}$ (see Fig.~\ref{fig:gamma_gb}).

\begin{figure}[h]
\includegraphics[width=\columnwidth]{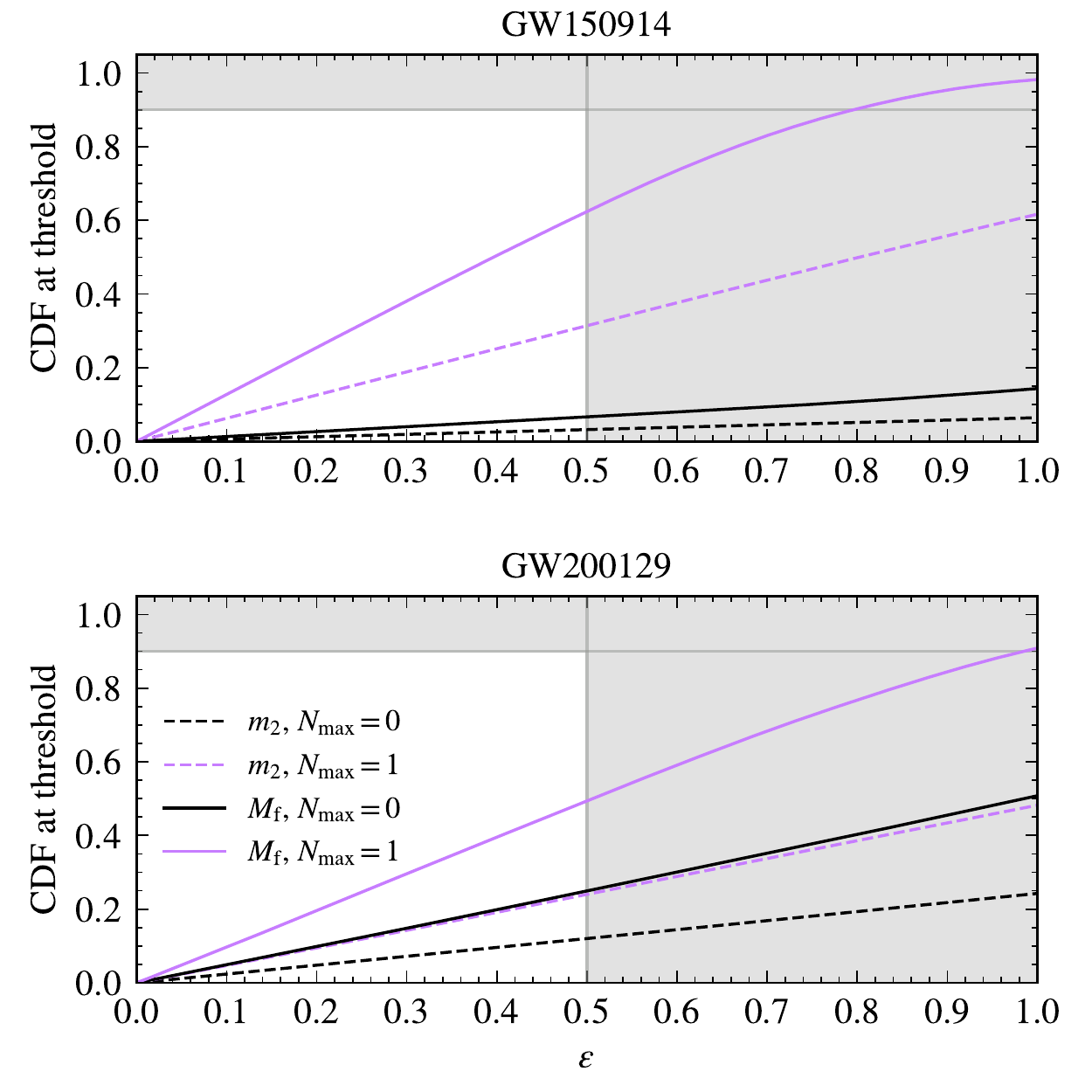}
\caption{The CDF evaluated at the
cutoff $\Lambda_{\rm EFT}(\varepsilon,\gm)$ for EdGB gravity as a function of the parameter
$\varepsilon$ for both (dashed curves) $\gm = m_2$, the secondary's source
mass, and (solid curves) $\gm = M_{\rm f}$, the remnant's source mass, without
(black curves) and with (purple curves) linear in spin QNM corrections. The
horizontal lines mark the $90\%$ credible levels. Having set the maximum value
of $\varepsilon$ to be 1/2, we see that in no situation the curves pass
through the 0.90 lines. This means that no bound on $\ell_{\rm EdGB}$ can be placed
with the events we analyzed.
}
\label{fig:edgb_cdf}
\end{figure}

As we have emphasized in Sec.~\ref{sec:remarks}, we must first check whether the
``EFT''~\eqref{eq:eft_bound} and ``ParSpec''~\eqref{eq:parspec_bound} bounds
are satisfied, before drawing any conclusions on the allowed values for
$\ell_{\rm EdGB}$ from our parameter estimations.
We check the validity of the EFT bound in Fig.~\ref{fig:edgb_cdf}. In the top (bottom) panel we show
the CDF of the $\ell_{\rm EdGB}$ posteriors for GW150914 (GW200129), obtained
by evaluating the integral~\eqref{eq:def_cdf} with $\ell^{\rm max}_{\rm th} = \Lambda_{\rm EFT}(\varepsilon, \mathfrak{m})$,
with the mass scale set by the secondary's mass (i.e., $\mathfrak{m} = m_2$, dashed lines) or
the remnant's mass ($\mathfrak{m} = M_{\rm f}$, solid lines), while varying $\varepsilon$ between 0 and 1.
For GW150914, we see that for the $N_{\rm max} = 0$ curves, that the CDF never goes
past $0.2$, regardless of the mass scale $\mathfrak{m}$ used and even at $\varepsilon =
1$, at which the EFT description of the theory would not be valid anyway.
This shows that the ``EFT bound'' given by Eq.~\eqref{eq:eft_bound} is
never met to a significant credible level and thus that we cannot place a bound on
$\ell_{\rm EdGB}$.
The situation is similar for GW200129 with $N_{\rm max} = 0$ and does not
change for either event when we add spin corrections to the EdGB QNM.
For the case with $N_{\rm max} = 1$, we find that the ``EFT bound'' is satisfied
only for $\varepsilon \approx 0.8$ and $\approx1$ for GW150914 and GW200129, respectively.
However, we set the maximum value of $\varepsilon$ to be 1/2, thus,
taken together we are led to \emph{conclude that we cannot constrain EdGB gravity with our
present model}.
We summarize our findings in Table~\ref{tab:summary_edgb}.
\begin{table}[h]
\begin{tabular}{l l c c c}
\hline
\hline
$N_{\rm max}$ & Event &  EFT    & ParSpec & Constraint      \\
              &       &  bound? & bound?  & ($\gm = M_{f}$) \\
\hline
  & GW150914 & No  & Yes & -- \\
0 & GW200129 & No  & Yes & -- \\
  & Combined & --  & Yes & -- \\
\hline
  & GW150914 & No  & Yes & -- \\
1 & GW200129 & No  & Yes & -- \\
  & Combined & --  & Yes & -- \\
\hline
\hline
\end{tabular}
\caption{Detailed summary of our results for EdGB gravity for GW150914, GW200129, and
combined events using $\gm = M_{\rm f}$, $\varepsilon = 1/2$ and quoting only 90\% credible results.
We find that we cannot place any bound on $\ell_{\rm EdGB}$ with our waveform model from either GW event.
}
\label{tab:summary_edgb}
\end{table}

We can compare this conclusion with that of Ref.~\cite{Carullo:2021dui},
which found that $p=4$ modifications (such as the case of EdGB gravity)
are constrained to $\ell \lesssim 35$~km,
\emph{but not} including theory-specific QNM information on $\delta \omega^{(j)}$ and $\delta \tau^{(j)}$.
Furthermore, Ref.~\cite{Carullo:2021dui} did not impose the EFT bound that we imposed.
Our results provide a concrete example of the importance of including
theory-specific QNM calculations information into the parameter estimation
and how this can dramatically change the outcome of the results.

Let us also contrast our results with those of Refs.~\cite{Nair:2019iur,Perkins:2021mhb,Lyu:2022gdr}
which relied on the BBH inspiral to constrain $\ell_{\rm EdGB}$, as discussed in Sec.~\ref{sec:review_edgb}.
We see that EdGB gravity provides an example of a theory in which, with current GW events, the inspiral portion
of the signal can be more constraining than the ringdown portion of the signal.
Two reasons together can explain our negative results. First, as observed by
Ref.~\cite{Blazquez-Salcedo:2016enn}, the QNMs of EdGB BHs only differ slightly
from their Schwarzschild counterparts. Second, the larger mass $M_{\rm f}$ of the
remnant BH, suppresses scalar field's charge relatively to the initial binary
components.

\subsection{Dynamical Chern-Simons gravity}
\label{sec:results_dcs}

\begin{figure}[t]
\includegraphics[width=\columnwidth]{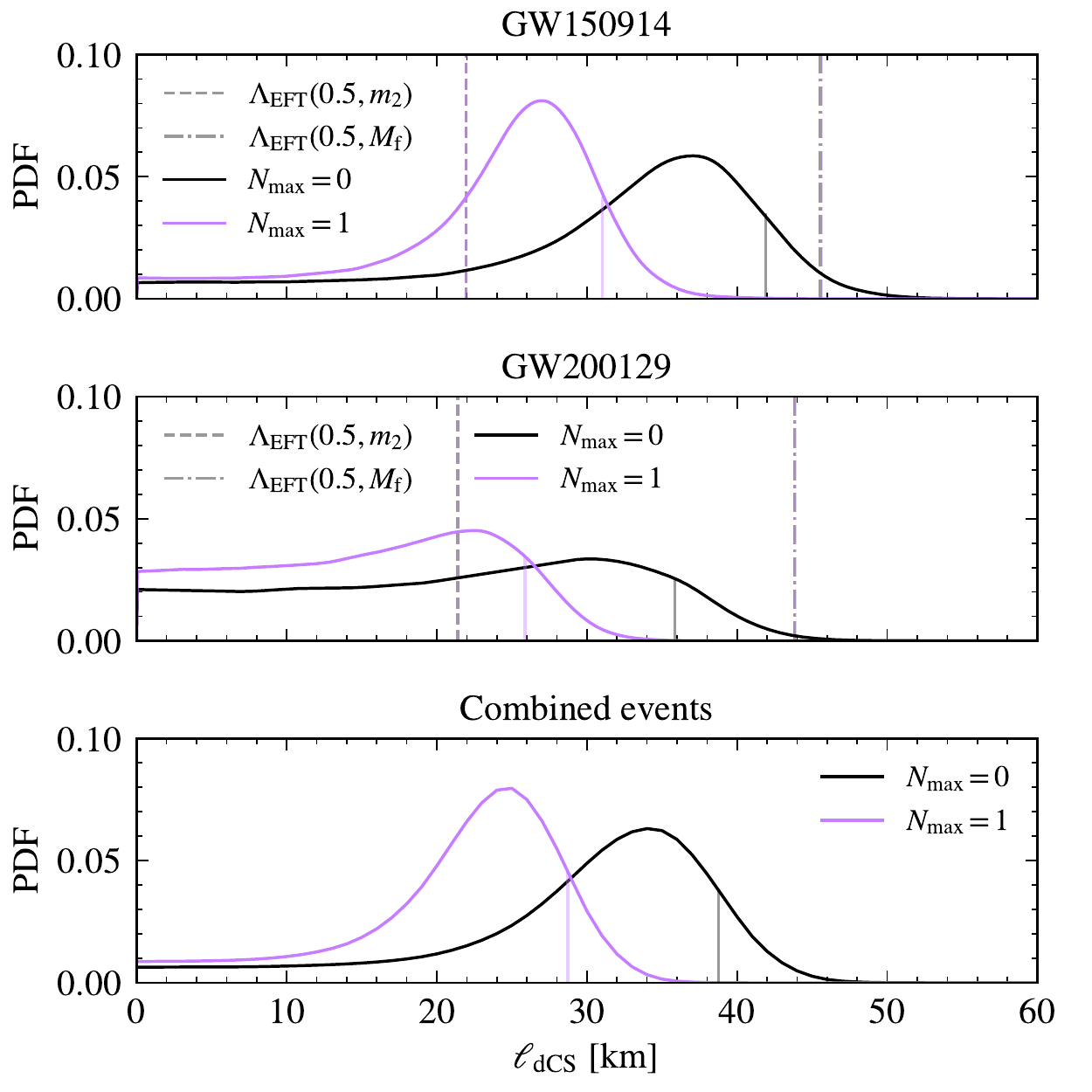}
\caption{Similar to Fig.~\ref{fig:edgb_exec_sum}, but for dCS gravity.
We stress that posteriors obtained including $n=1$ corrections, violate conditions~\eqref{eq:parspec_bound}
and therefore should not be used to draw meaningful conclusions. We show them for illustrative purposes
and also to emphasize the importance of taking conditions~\eqref{eq:eft_bound} and~\eqref{eq:parspec_bound}
simultaneously into consideration when analyzing the results of the parameter estimation.
}
\label{fig:dCS_exec_sum}
\end{figure}

\begin{figure}[t]
\includegraphics[width=\columnwidth]{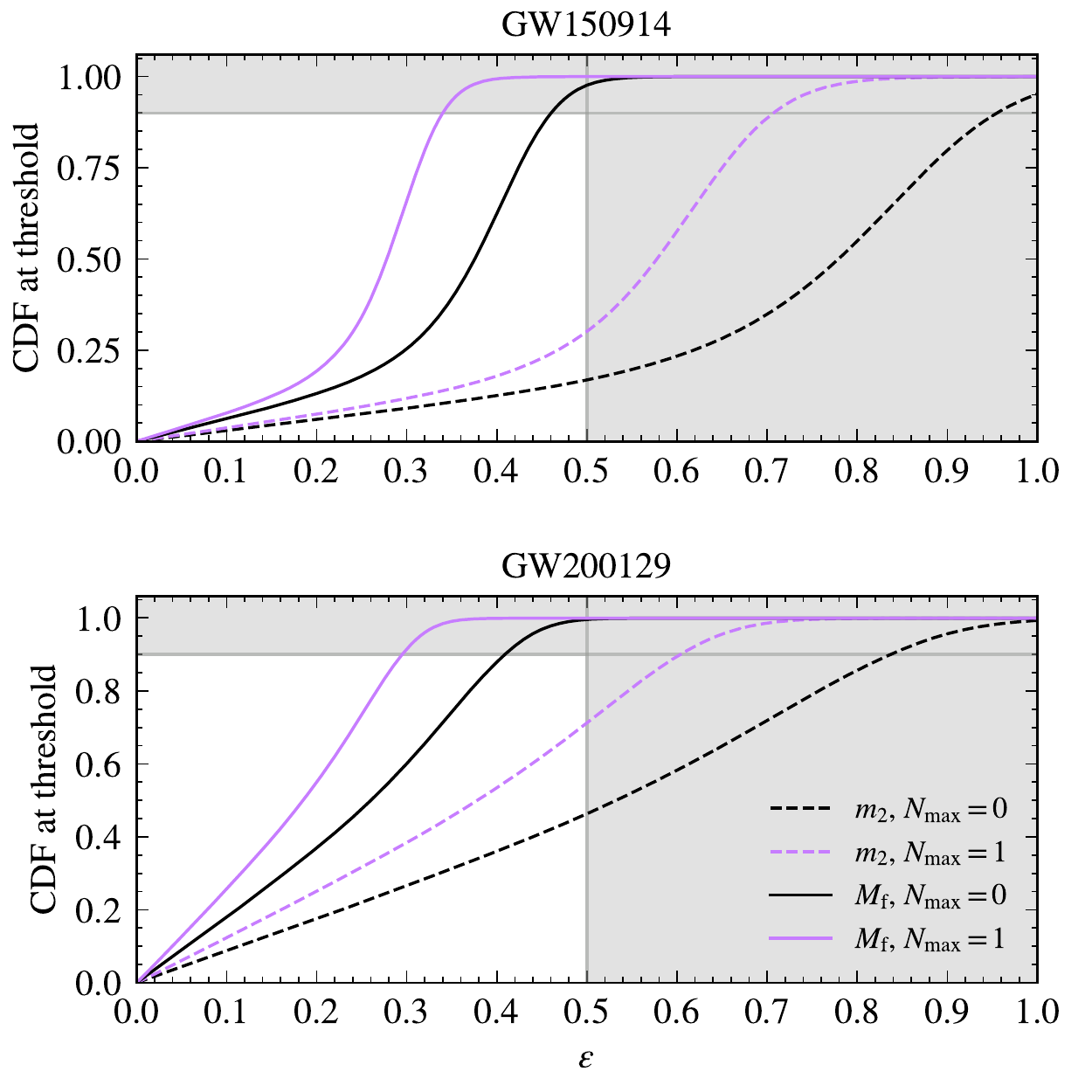}
\caption{Similar to Fig.~\ref{fig:edgb_cdf}, but for dCS gravity.
We see that the CDF curves for $\gm = M_{\rm f}$ are above 90\% for $\varepsilon =
1/2$ for both events, with and without including the $n = 1$ dCS corrections to
the dominant QNM.
}
\label{fig:dcs_cdf}
\end{figure}

We now consider dCS gravity where the main results are summarized in Fig.~\ref{fig:dCS_exec_sum}.
As with EdGB gravity (see Sec.~\ref{sec:results_edgb}) although the PDF of $\ell_{\rm dCS}$
is peaked away from 0, this does not signify a deviation from GR, as we have verified that,
\begin{equation}
    \gamma_{\rm CS} = \left( \frac{c^2 \ell_{\rm dCS}}{G M^{\rm s}_{\rm f}} \right)^{4}\,,
\end{equation}
does indeed peak at zero indicating consistency with GR, similarly to what is
shown in Fig.~\ref{fig:gamma_gb} for $\gamma_{\rm EdGB}$.
We also see that in both cases the inclusion of leading-order--in-spin
correction to the QNM displaces the posteriors toward smaller values of
$\ell_{\rm dCS}$. This can be seen more evidently by looking at the location of
posterior peaks.
Finally, in the bottom panel, we show the combined result for both events.

In Fig.~\ref{fig:dcs_cdf} we show the CDF for GW150914, we see that with $\gm = m_2$,
Eq.~\eqref{eq:eft_bound} is not satisfied unless $\varepsilon \approx 0.9$ (with only $j=0$ corrections) and
$\varepsilon \approx 0.7$ (with both $j=0$ and $1$ corrections).
The situation is different if we use $\gm = M_{\rm f}$. In this case, we find that
with or without spin corrections Eq.~\eqref{eq:eft_bound} can be satisfied
with $\varepsilon \leqslant 1/2$ (i.e., below the criteria used
Refs.~\cite{Nair:2019iur,Perkins:2021mhb,Lyu:2022gdr}).
This means that with our model's assumptions and using the remnant's source mass $M_{\rm f}$ to set the
cutoff scale that we can claim an upper bound
\begin{equation}
\ell_{\rm dCS} \leqslant 41.9~\textrm{km}
\quad \textrm{at 90\% credible level},
\end{equation}
on dCS gravity. This result would constitute \emph{the strongest bound to date on this
theory with GW observations alone, and also the first bound using GW generation effects.}

We can draw qualitatively similar conclusions from the GW200129 event.
In particular, we find,
\begin{equation}
\ell_{\rm dCS} \leqslant 35.8~\textrm{km}
\quad \textrm{at 90\% credible level}.
\end{equation}
These stronger bounds are a consequence of the larger support for $\ell_{\rm dCS} \lessapprox 15$~km
for GW200129 (compare the top and middle panels in Fig.~\ref{fig:dCS_exec_sum}), and in part due to
the smaller median remnant ($M_{\rm f} \approx 59.5\msun$ versus $M_{\rm f} \approx 61.8\msun$ for GW150914).
We also found for both GW events, that the perturbative-conditions~\eqref{eq:parspec_bound} required by the ParSpec
is violated for the $\gamma_{\rm dCS} \, \delta \tau^{(1)}$ coefficient.
This means that we cannot use this posterior to infer any meaningful bound on dCS gravity
and that is why we quoted only the $N_{\rm max}=0$ bound above.

Finally, since both events individually lead to a bound on $\ell_{\rm dCS}$
(assuming a cutoff scale for $\gm = M_{\rm f}$ and $N_{\rm max}=0$), we can combine the
posteriors to obtain the cumulative bound,
\begin{equation}
\ell_{\rm dCS} \leqslant 38.7~\textrm{km}
\quad \textrm{ at 90\% credible level},
\end{equation}
which is the main result of this section.
This bound is approximately a factor of four weaker than that placed by
Ref.~\cite{Silva:2020acr}, but it
(i) relies only on GW observations, and
(ii) suggests that a ringdown analysis can potentially place constraints on
theories that, with current GW events, can evade GR tests using inspiral information
alone, such as the case of dCS gravity~\cite{Nair:2019iur,Perkins:2021mhb,Lyu:2022gdr}.
In Table~\ref{tab:summary_dcs} we summarize our findings of this section.

\begin{figure}[t]
\includegraphics[width=0.9\columnwidth]{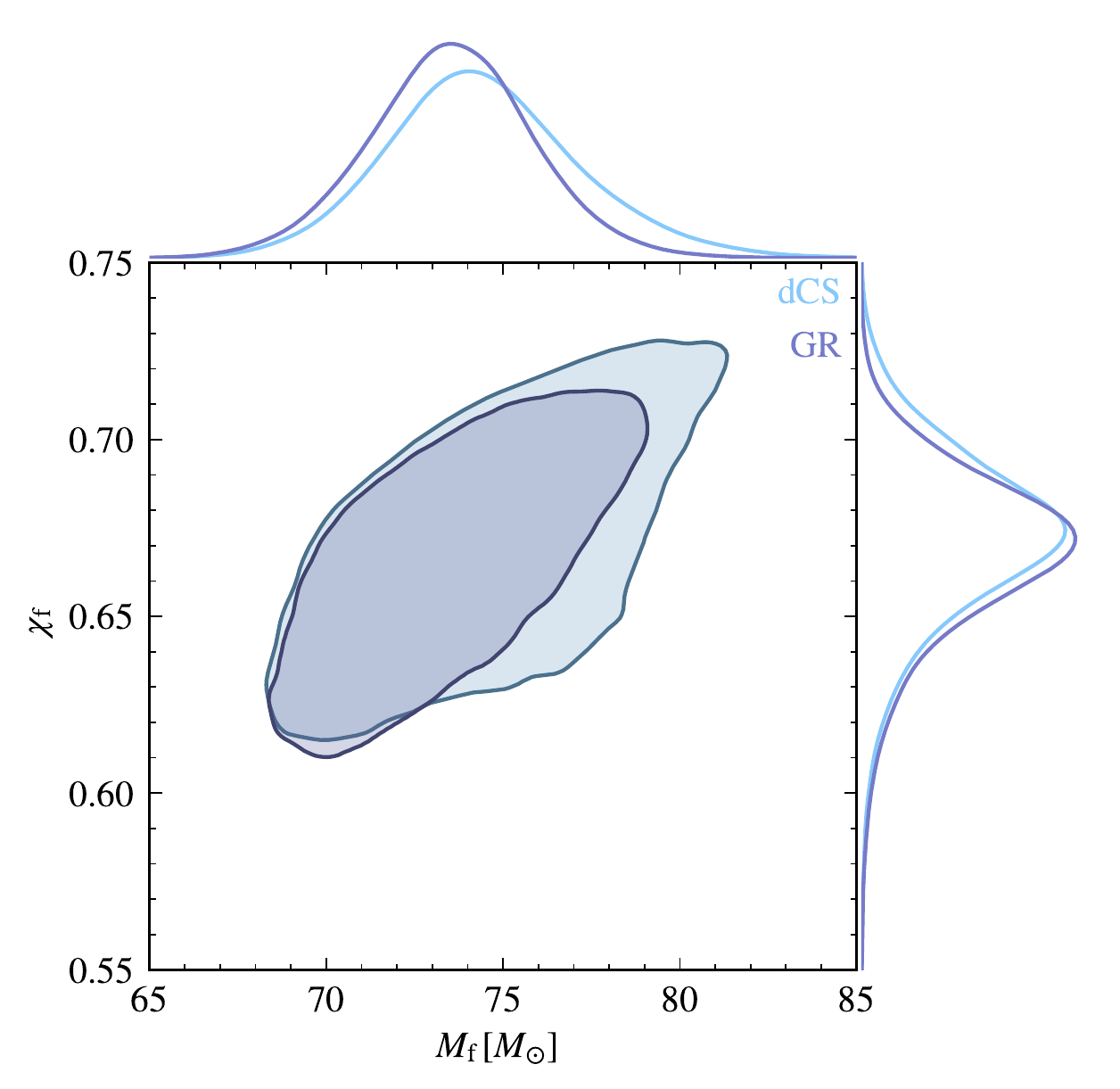}
\caption{Corner plot showing that the inferred final spin $\chi_{\rm f}$,
remnant mass $M_{\rm f}$ for GW150914, using the same waveform model,
but without (purple contours) and with the non-GR parameters
different from zero (blue contours), for
dCS gravity and $N_{\rm max}=0$.
The contours represent 90\% credible levels.
We see that the introduction of the non-GR parameters does not bias the
inference on the source parameters as required by the ParSpec.
}
\label{fig:corner_plot}
\end{figure}

\begin{table}[h]
\begin{tabular}{l l c c c}
\hline
\hline
$N_{\rm max}$ & Event & EFT    & ParSpec & Constraint ($\gm = M_{f}$) \\
              &       & bound? & bound?  &                            \\
\hline
      & GW150914 & Yes & Yes & $\ell_{\rm dCS} \leqslant 41.9$~km \\
0     & GW200129 & Yes & Yes & $\ell_{\rm dCS} \leqslant 35.8$~km \\
      & Combined &     &     & \cellcolor{black!10}$\ell_{\rm dCS} \leqslant 38.7$~km \\
\hline
      & GW150914 & Yes & No  & --                                 \\
1     & GW200129 & Yes & No  & --                                 \\
      & Combined &     &     & --                                 \\
\hline
\hline
\end{tabular}
\caption{Detailed summary of our results for dCS gravity for GW150914, GW200129, and
combined events using $\gm = M_{\rm f}$, $\varepsilon = 1/2$ and quoting only 90\% credible bounds.
We found that while our posteriors satisfy the condition~\eqref{eq:eft_bound} (with $\varepsilon = 1/2$),
they do not obey the condition~\eqref{eq:parspec_bound} for $N_{\rm max} = 1$. This means
that our results for $N_{\rm max}=0$ are the only ones we can confidently quote.
The combined bound, which is also quoted in Table~\ref{tab:ref_theories_qnms},
is $\ell_{\rm dCS} \leqslant 38.7$~km at 90\% credible level.
}
\label{tab:summary_dcs}
\end{table}

As an additional check, to verify the robustness of our constraint, we show in
Fig.~\ref{fig:corner_plot}, the final spin $\chi_{\rm f}$ and remnant mass
$M_{\rm f}$ for GW150914 for GR and dCS gravity.
We see that our \pSEOB~waveform model does not introduce substantial changes
to the GR estimates on these parameters, as required by the ParSpec expansion
(see discussion in Sec.~\ref{sec:review_parspec}).
In fact, we observed no bias on the estimation of $\Mf$ and $\chi_{\rm f}$
for all theories considered here.
For completeness, in Appendix~\ref{app:big_contour} we also
show how all other intrinsic parameters remain unbiased.

\subsection{Cubic effective-field-theory of general relativity}
\label{sec:results_ceft}

\begin{figure}[t]
\includegraphics[width=\columnwidth]{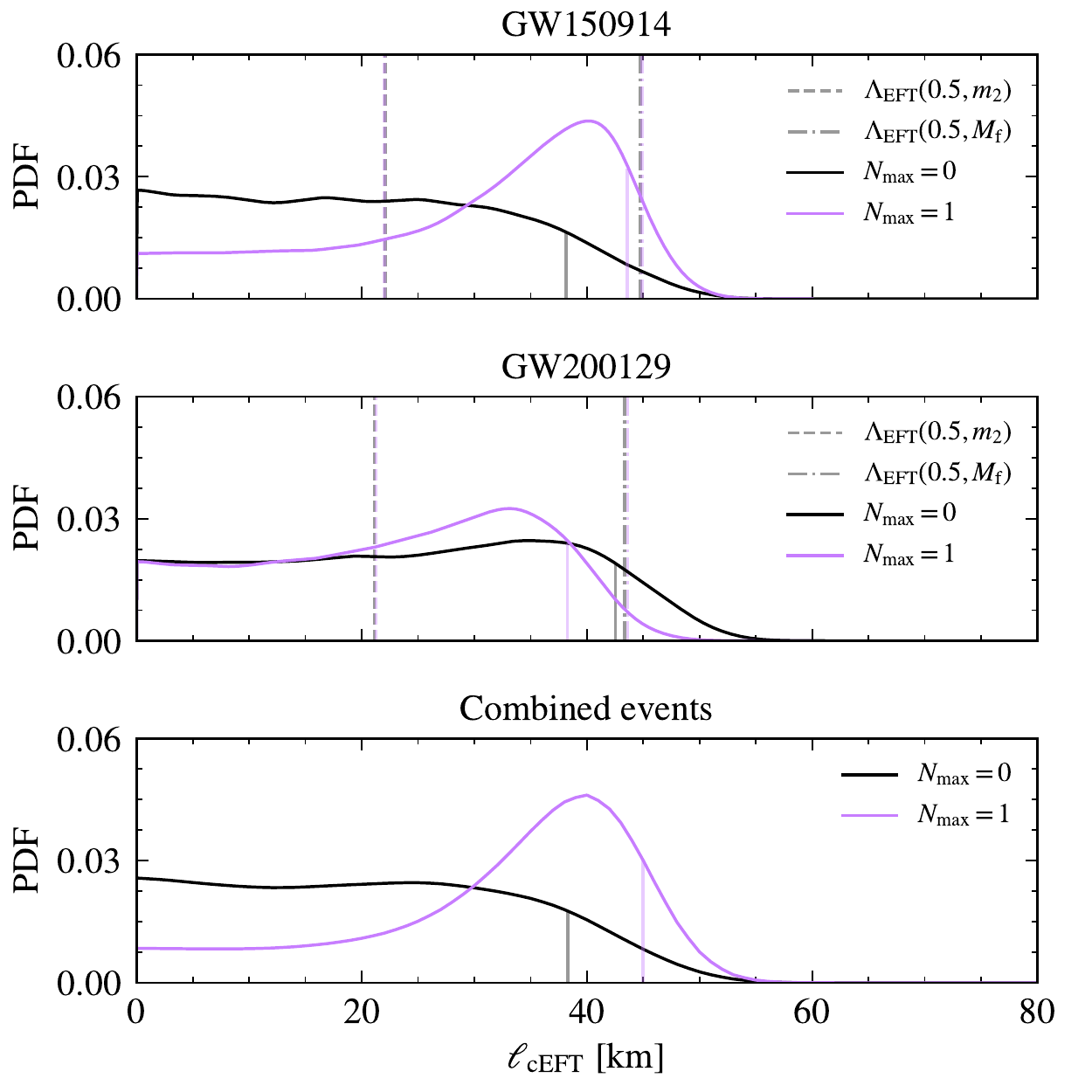}
\caption{Similar to Fig.~\ref{fig:edgb_exec_sum}, but for the cubic EFT of GR.
We show our results for
GW150914 (top panel), GW200129 (middle panel) and combined events (bottom panel).
The colors distinguish different $N_{\rm max}$ in the spin expansion. Once again, the
solid vertical lines mark the 90\% upper credible intervals, while
the dashed and dot-dashed lines correspond to the EFT bound, $\Lambda_{\rm EFT}(0.5, m_2)$ and
$\Lambda_{\rm EFT}(0.5, M_{\rm f})$, respectively.
}
\label{fig:cEFT_exec_sum}
\end{figure}

We now consider the cubic EFT of GR.
In Fig.~\ref{fig:cEFT_exec_sum} we show the marginalized posterior distributions
functions of $\ell_{\rm cEFT}$ for GW150914 (top panel) and GW200129 (middle panel),
with different curve colors corresponding to different $N_{\rm max}$ in
the spin expansion.
We find that in this theory, the posterior distributions are mostly uniform for
$\ell_{\rm cEFT} \lesssim 40$~km (contrast this with the EdGB and dCS gravity
cases in Figs.~\ref{fig:edgb_exec_sum} and~\ref{fig:dCS_exec_sum}).
For values $\ell_{\rm cEFT} \gtrsim 40$~km, the posteriors smoothly approach zero.

\begin{figure}[h]
\includegraphics[width=\columnwidth]{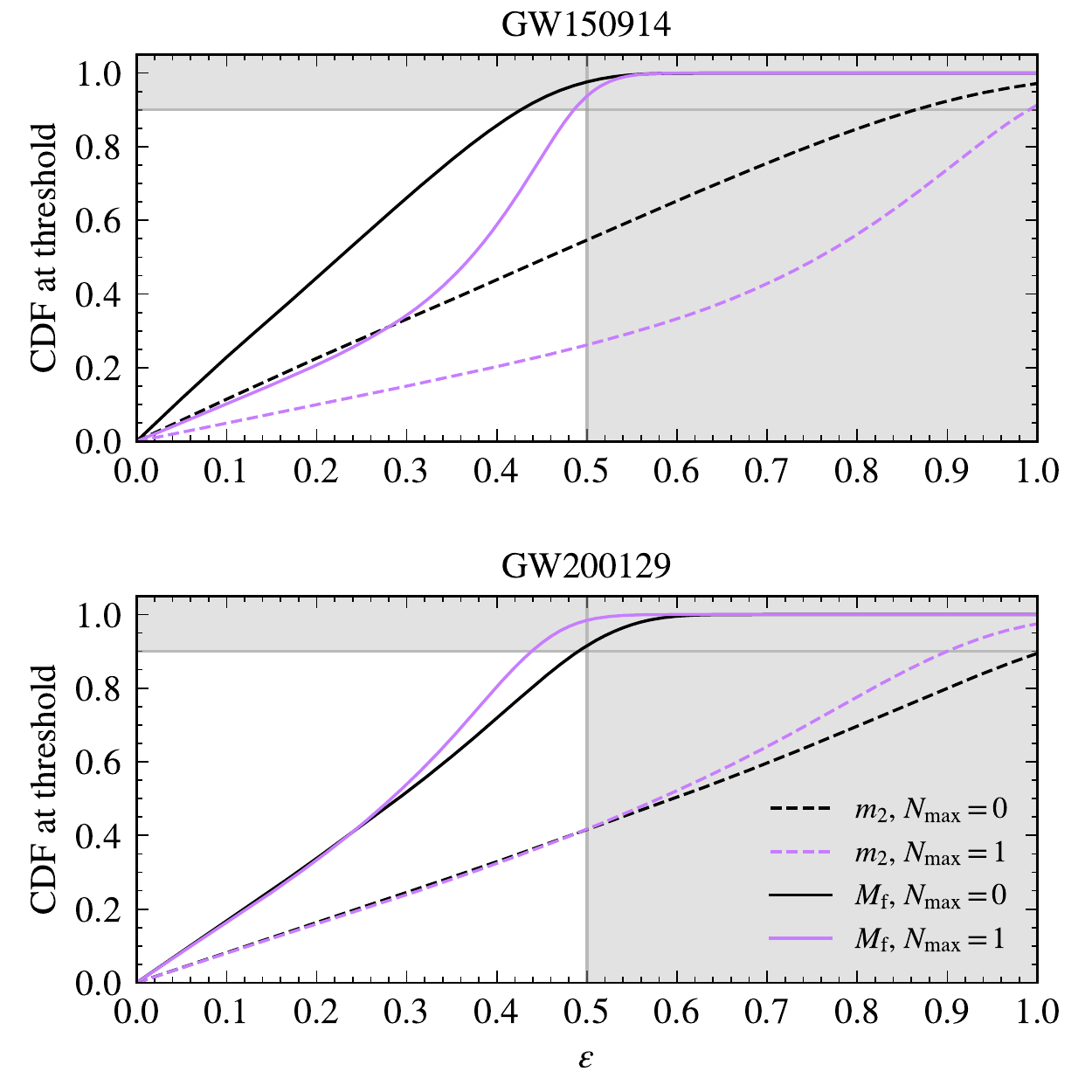}
\caption{Similar to Fig.~\ref{fig:edgb_cdf}, but for the cubic EFT of GR.
We see that the CDF curves for $\gm = M_{\rm f}$ are above 90\% for $\varepsilon =
1/2$ for both events, with and without including the $j = 1$ corrections to
the dominant QNM.
}
\label{fig:cEFT_cdf}
\end{figure}

In Fig.~\ref{fig:cEFT_cdf} we show the CDF for both events, calculated in the
same way as already described for the EdGB and dCS theories.
We see that curves are very similar to those of dCS gravity for GW200129 (see
bottom panel in Fig.~\ref{fig:cEFT_cdf}).
Moreover, we find that the EFT~\eqref{eq:eft_bound} and
ParSpec~\eqref{eq:parspec_bound} bounds are satisfied for both events both when
$\mathfrak{m} = M_{\rm f}$, $\varepsilon = 1/2$, and $N_{\rm max} = 0$.
This allows us to place the combined bound of
\begin{equation}
    \ell_{\rm cEFT} \leqslant 38.2~\textrm{km}, \quad \textrm{at 90\% credible level}\,.
\end{equation}
As also happened for our study for dCS, the find that, for
the cubic EFT, the ParSpec bound is violated by the $N_{\rm max} = 1$ corrections to
the QNMs, meaning that we cannot use this case to draw any meaningful
constraint on this parameter.
We summarize our results in Table~\ref{tab:summary_ceft}.
\begin{table}[h]
\begin{tabular}{l l c c c}
\hline
\hline
$N_{\rm max}$ & Event & EFT    & ParSpec & Constraint ($\gm = M_{f}$) \\
              &       & bound? & bound?  &                            \\
\hline
  & GW150914  & Yes & Yes &  $\ell_{\rm cEFT} \leqslant 38.2$~km \\
0 & GW200129  & Yes & Yes &  $\ell_{\rm cEFT} \leqslant 42.5$~km \\
  & Combined  &     &     &  \cellcolor{black!10}$\ell_{\rm cEFT} \leqslant 38.2$~km \\
\hline
  & GW150914  & Yes  & No  &  -- \\
1 & GW200129  & Yes  & No  &  -- \\
  & Combined  &      &     &  -- \\
\hline
\hline
\end{tabular}
\caption{Detailed summary of our results the cubic EFT of GR for GW150914, GW200129, and
combined events using $\gm = M_{\rm f}$, $\varepsilon = 1/2$ and quoting only 90\% credible results.
}
\label{tab:summary_ceft}
\end{table}


\subsection{Quartic effective-field-theory of general relativity}
\label{sec:results_qeft}

\begin{figure}[t]
\includegraphics[width=\columnwidth]{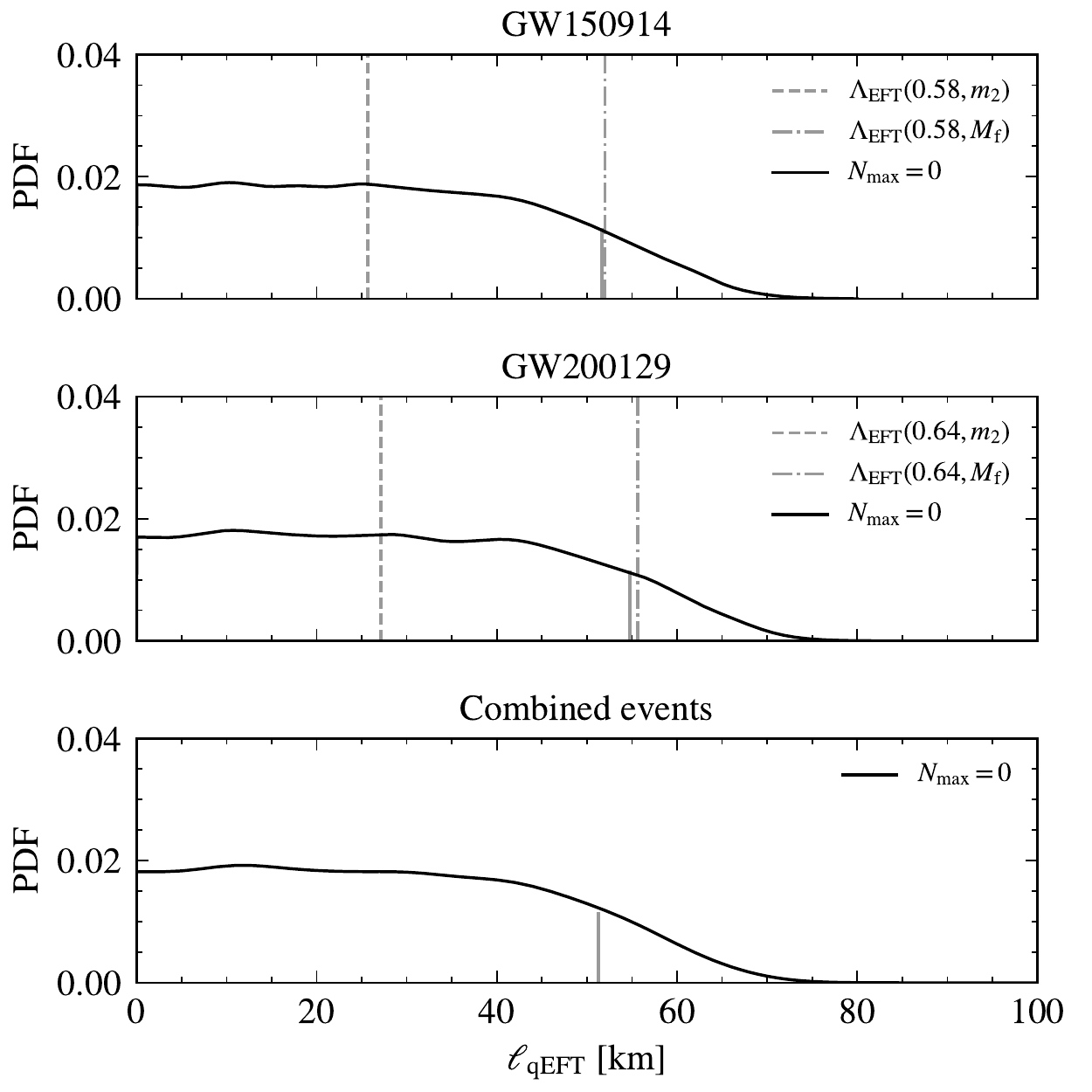}
\caption{Similar to Fig.~\ref{fig:edgb_exec_sum}, but for the quartic EFT of GR.
We show our results for GW150914 (top panel), GW200129 (middle panel) and combined events (bottom panel).
As in Fig.~\ref{fig:edgb_exec_sum}, we mark with the solid vertical lines the 90\% upper credible intervals.
Here we calculated the values of $\Lambda_{\rm EFT}(\varepsilon, \mathfrak{m})$ using
$\varepsilon = 0.58$ (for GW150914) and $\varepsilon = 0.64$ (for GW200129) as discussed in
main text.
}
\label{fig:qEFT_exec_sum}
\end{figure}

\begin{figure}[t]
\includegraphics[width=\columnwidth]{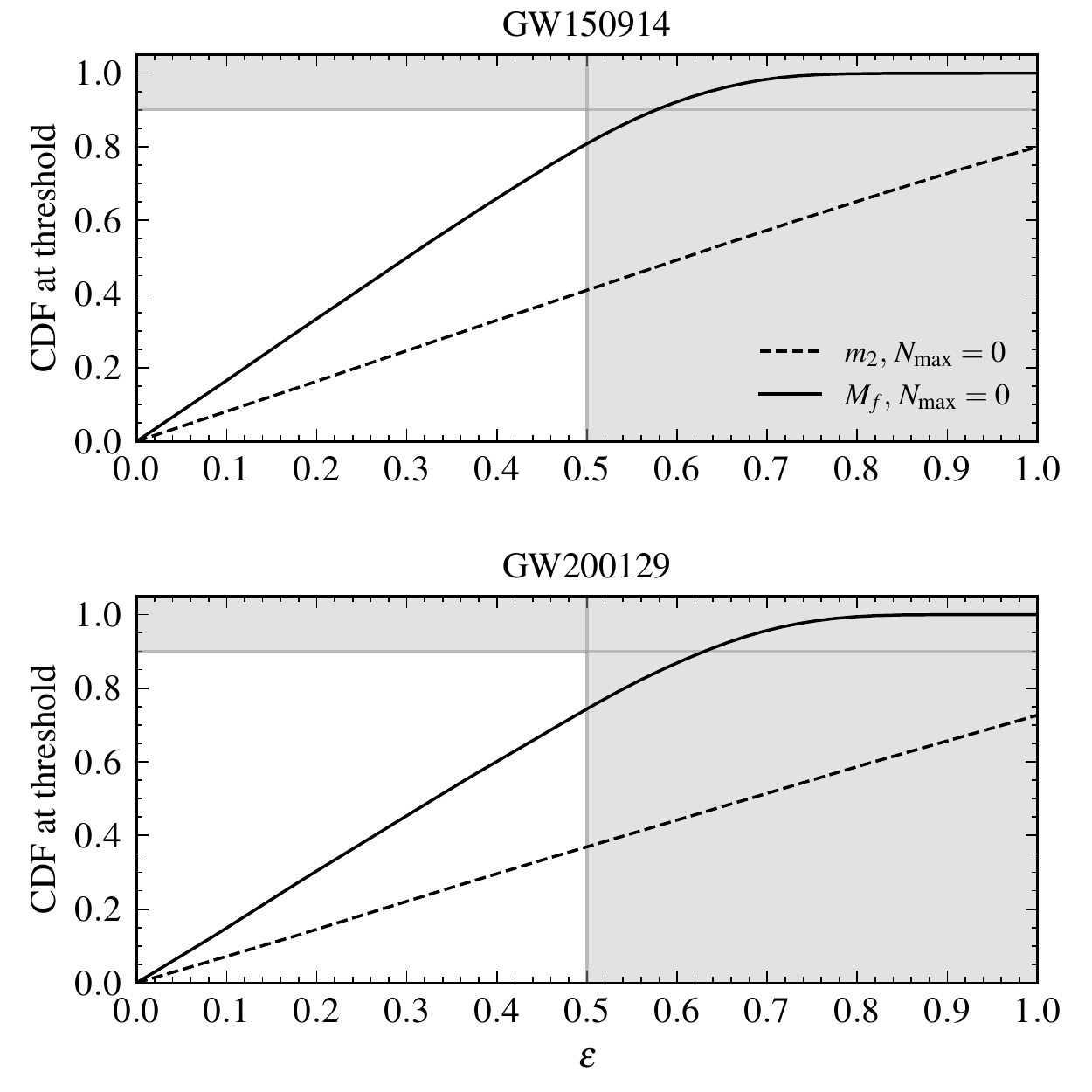}
\caption{Similar to Fig.~\ref{fig:edgb_cdf}, but for the quartic EFT of GR.
We find that similarly to what happens in dCS gravity and the cubic EFT of GR, we
can place a constraint on $\ell_{\rm qEFT}$ when $\gm = M_{\rm f}$ for both events at 68\% credibility.
}
\label{fig:qEFT_cdf}
\end{figure}

Let us now consider the quartic EFT of GR, as our final example.
In Fig.~\ref{fig:qEFT_exec_sum} we show the posteriors on $\ell_{\rm qEFT}$ for GW150914 (top panel),
GW200129 (middle panel) for $N_{\rm max} = 0$, which are qualitatively similar to the cubic EFT of GR.
We find that while the Parspec bound is satisfied, the EFT bound is only marginally so,
As shown in Fig.~\ref{fig:qEFT_cdf}, the 90\% credible level is reached for $\varepsilon \approx 0.58$
(in the case of GW15094) and for $\varepsilon \approx 0.64$ (in the case of GW200129).
Having in mind that the cut off $\varepsilon = 1/2$ is not fundamental, but to keep consistency
across our analysis, our final result
\begin{equation}
    \ell_{\rm qEFT} \leqslant 51.3~\textrm{km},
\end{equation}
at 90\% credible level should be taken lightly.
However, we \emph{can} claim the validity of the bound above, but at a lower,
68\% credible level.

\begin{table}[b]
\begin{tabular}{l l c c c}
\hline
\hline
$N_{\rm max}$ & Event & EFT    & ParSpec & Constraint ($\gm = M_{\rm f}$) \\
              &       & bound? & bound?  &                            \\
\hline
  & GW150914 & Yes & Yes  & $\ell_{\rm qEFT} \leqslant 51.7$~km \\
0 & GW200129 & Yes & Yes  & $\ell_{\rm qEFT} \leqslant 54.8$~km \\
  & Combined &     &      & \cellcolor{black!10}$\ell_{\rm qEFT} \leqslant 51.3$~km \\
\hline
\hline
\end{tabular}
\caption{Detailed summary of our results the quartic EFT of GR for GW150914, GW200129, and
combined events using $\gm = M_{\rm f}$, $\varepsilon = 1/2$, and $N_{\rm max} = 0$. The quoted
result correspond to 90\% credible values \emph{if} we allow for a more flexible cutoff $\varepsilon \lesssim 0.65$.
However, the result is robust for the cut off $\varepsilon = 1/2$, at 65\% credible level.
}
\label{tab:summary_qeft}
\end{table}

In this theory, we have considered only $N_{\rm max} = 0$.
We find that the addition of spin corrections (while maintaining the same prior
ranges on $\ell_{\rm qEFT}$ as used in the $N_{\rm max} = 0$ study) can result
in waveforms that can have a ringdown segment larger (sometimes seconds long)
than the inspiral-plunge segment in the detectors' frequency band, making the
parameter estimation challenging.
To overcome this issue we have lowered the value of $\ell^{\rm max}_{\rm qEFT}$, but by doing so we have obtained posteriors
which were flat, just as our prior, and were thus uninformative. Hence, we do not quote any results for $N_{\rm max} = 1$.
Table~\ref{tab:summary_qeft} summarizes our findings for the quartic EFT of GR.

\section{Conclusions}
\label{sec:conclusions}

We presented an unified framework that combines the ParSpec framework
to model deviations to the GR QNMs~\cite{Maselli:2019mjd} with the \pSEOB~waveform model~\cite{Brito:2018rfr,Ghosh:2021mrv}.
We showed with concrete examples, how theory-specific QNM calculations of
slowly rotating BHs in modified gravity theories can be mapped onto the
non-GR parameters of the ParSpec formalism.
The resulting \pSEOB~waveform model does not bias (relative to GR) the inference of
the intrinsic binary parameters, as required by ParSpec (see, in particular, Fig.~\ref{fig:corner_plot} and
Fig.~\ref{fig:corner_plot_all} in Appendix~\ref{app:big_contour}),
Put together this allowed us to test four modified gravity theories (EdGB, dCS,
cubic, and quartic EFTs of GR) using observational data from the LVK events
GW150914 and GW200129. Our results are summarized in Table~\ref{tab:bound_summary}.

In particular, we found, that within the interpretation of these theories as EFTs and the
region of validity of the ParSpec framework, the fundamental lengthscale of dCS gravity is bound as
$\ell_{\rm dCS} \leqslant 34.5$~km, at 90\% credible level, when stacking the posteriors
of GW150914 and GW200129. This is the strongest constrain to date on this
theory with GW observations alone.
In contrast, we could not place any bounds on the fundamental lengthscale
of EdGB gravity $\ell_{\rm EdGB}$.
This dichotomy between the two theories has a counterpart with works that considered
the inspiral part of the GW signal alone~\cite{Nair:2019iur,Perkins:2021mhb,Lyu:2022gdr}.
Using data of the LVK BBHs, it was found that the posterior distributions for
deviations from GR were uninformative in dCS gravity, but not in EdGB gravity.
We emphasize that both those theories (and the cubic EFT of GR also studied here) all predict the same exponent $p$ in ParSpec.
Hence, our results show how the inclusion of theory-specific information into the ParSpec framework
can result in different outcomes for different theories, even if they predict the same value of $p$.

Let us discuss some avenues for future work.
First, we could implement a high-spin version of the GR fitting coefficients to the ParSpec formulas.
This has already been done in Ref.~\cite{Carullo:2021dui} extending the validity of the ParSpec formulas up to
spins of $\chi_{\rm f} \approx 0.99$.
For the events analyzed here, the original fit by Ref.~\cite{Maselli:2019mjd}
was sufficient, but it might not be the case with upcoming GW observation campaigns.
Second, it would be important to incorporate additional effects, such as spin-precession and eccentricity
to \pSEOB~(see, for instance, Refs.~\cite{Ossokine:2020kjp,Ramos-Buades:2021adz}).
Third, it will be
interesting, to perform tests of modified theories of gravity using IMR waveform models that include, during
the inspiral stage, finite-size effects induced by the non-GR geometry around the BHs ---
for example the ones due to spin-induced quadrupole, tidal deformability and absorption, and also
orbital effects due to non-GR gravitational interactions
between the BHs. Those effects could be included using the flexible theory-independent
method~\cite{Mehta:2022pcn}, as done in Ref.~\cite{Sennett:2019bpc} or the
TIGER code~\cite{Li:2011cg,Agathos:2013upa}. However, setting bounds on deviations
from GR caused by orbital effects requires a different EFT interpretation than what we
adopted in Sec.~\ref{sec:remarks} (see also Sec.~IIC in Ref.~\cite{Sennett:2019bpc}).
Indeed, in this case one would need to analyze the data considering that the modified theory of
gravity is valid for $\ell_{\rm th} \gtrsim  M$, but $\ell_{\rm th} \lesssim \mathcal{R}$, being $\mathcal{R}$ the binary's separation.
Fourth, to test the robustness of the results obtained in this paper, it will be very useful to employ NR waveforms
produced in some of the non-GR theories under consideration, as synthetic signals, and carry out
Bayesian analysis to recover the binary's parameters, including the non-GR ones during the ringdown.
As today, there are only a small number of such BBH NR simulations, for a
given theory~\cite{Healy:2011ef,Witek:2018dmd,Okounkova:2017yby,Hirschmann:2017psw,Okounkova:2020rqw,Okounkova:2020rqw,Silva:2020omi,East:2020hgw,East:2021bqk,Figueras:2021abd}.
Last, but not least, it will be very
beneficial to calculate the QNMs (complex) frequencies of rapidly rotating BHs in modified gravity theories
(with BH perturbation theory). This is a challenging problem, but certainly necessary,
also in the context of ParSpec, if higher spin corrections would need to be
included to make the framework robust.

\section*{Acknowledgments}
\label{sec:acknowledgements}
We thank Emanuele Berti, Andrea Maselli, Caio F. B. Macedo, Deyan Mihaylov,
Serguei Ossokine, Scott E. Perkins, and Helvi Witek for discussions.
We also thank Anuradha Gupta and Gregorio Carullo for comments on this manuscript.
We also thank the referees for the detailed review of this work and for
the suggestions that improved its presentation.
We are grateful for the computational resources provided by the Max Planck
Institute for Gravitational Physics in Potsdam, specifically the
high-performace computing cluster Hypatia and to Steffen Grunewald for
assistance.
We acknowledge funding from the Deutsche Forschungsgemeinschaft (DFG) - project number: 386119226.
The material presented in this manuscript is based upon work supported by NSF’s LIGO Laboratory,
which is a major facility fully funded by the NSF.
This research has made use of data, software and/or web tools obtained from the Gravitational Wave Open
Science Center (https://www.gw-openscience.org), a service of LIGO Laboratory, the LIGO Scientific Collaboration
and the Virgo Collaboration, and Zenodo (https://zenodo.org/record/5172704).
The authors would like to thank everyone at the frontline of the Covid-19
pandemic.

\appendix

\section{Details of the determination of the theory-specific ParSpec coefficients}
\label{app:map_details}

Here, for the theories described in Sec.~\ref{sec:review_theories}, we
use QNM calculations from the literature and determine
the coefficients in the ParSpec, which we have summarized
in Table~\ref{tab:ref_theories_qnms}.
We consider only the fundamental QNM $(\ell, m, n) = (2, 2, 0)$,
hence we omit the QNM subscript ``$(2,2,0)$'' for brevity and, likewise,
the subscript ``f'' for final BH's spin and mass.

\subsection{Einstein-dilaton-Gauss-Bonnet gravity}
\label{app:map_edgb}

We start by considering EdGB gravity and focus on
Refs.~\cite{Blazquez-Salcedo:2016enn,Pierini:2021jxd} to determine the ParSpec coefficients for this theory.
In particular, Ref.~\cite{Blazquez-Salcedo:2016enn} found that the damping time
of the dominant \emph{axial} gravitational-led mode increases as the lengthscale
$\ell_{\rm EdGB}$ is increased. The leading-order spin corrections to the
\emph{polar}-parity QNMs was studied in Ref.~\cite{Pierini:2021jxd}.
Hence, according to the prescription of Sec.~\ref{sec:theory_specific_qnm},
we select the axial-parity branch of QNMs.
For the nonrotating QNMs we use
the numerical data of Ref.~\cite{Blazquez-Salcedo:2016enn} and
generate a new linear fit in $\gamma_{\rm EdGB}$ using numerical QNM data valid for small
values of the $\gamma_{\rm EdGB}$ (see, in particular, Eq.~(27) and Fig.~1 of
Ref.~\cite{Blazquez-Salcedo:2016enn}).
We find,
\begin{subequations}
\begin{align}
    M {\rm Re}(\sigma)_{\rm EdGB} &= M {\rm Re}(\sigma)_{\rm GR} \, \left( 1 + 0.0107 \, \gamma_{\rm EdGB} \right),
    \\
    M {\rm Im}(\sigma)_{\rm EdGB} &= M {\rm Im}(\sigma)_{\rm GR} \, \left( 1 - 0.0044 \, \gamma_{\rm EdGB} \right).
\end{align}
\end{subequations}
The small values of the numerical prefactors of $\gamma_{\rm EdGB}$ are a consequence of the how weakly the QNMs
of BHs in EdGB gravity deviate from their GR counterparts, even at moderately large values of $\gamma_{\rm EdGB} \approx 0.3$.

The spin-corrections to the \emph{polar} gravitational-led modes
were calculated in Ref.~\cite{Pierini:2021jxd} (see, in particular, their Eqs.~(51)
and~(52)).
For consistency with our previous discussion, we truncate these equations at
leading-order in $\gamma_{\rm EdGB}$, but we emphasize that we are being inconsistent in
mixing results valid for modes of different parities.
We still do so, simply to explore what the rotational corrections to EdGB gravity QNMs \emph{might}
tells us in our ringdown analysis and the results of Ref.~\cite{Pierini:2021jxd} are our best
presently available guide.

We can expand the resulting formula in $\gamma_{\rm EdGB}$ and the coefficients $\delta \omega^{(i)}$ and $\delta \tau^{(i)}$, $i=1,2$ can be
read-off by comparison against Eqs.~\eqref{eq:qnm_nongr_iso}, where for the damping time we use the relation ${\rm Im}(\sigma)_{\rm EdGB} = - 1 / \tau_{\rm EdGB}$
and reexpand in $\ell_{\rm EdGB}$ and $\chi$.
These steps yield for $p_{\rm EdGB} = 4$,
\begin{align}
     \delta \omega_{\rm EdGB}^{(0)} = 0.0107, \quad \delta \tau_{\rm EdGB}^{(0)} = -0.2480,
\end{align}
for the $j=0$ coefficients and
\begin{equation}
    \delta \omega_{\rm EdGB}^{(1)} = -0.2480, \quad \delta \tau_{\rm EdGB}^{(1)} = -1.1014.
\end{equation}
for the $j=1$ coefficients.

\subsection{Dynamical Chern-Simons gravity}
\label{app:map_dcs}

For dCS gravity, we follow Ref.~\cite{Wagle:2021tam}, which numerically calculated the QNMs
of slowly rotating BHs, and found that for the axial gravitational-led modes the damping time increases, as we
increase $\ell_{\rm dCS}$, at constant, small BH spin.
Hence, according to the prescription of Sec.~\ref{sec:theory_specific_qnm},
this is the branch of QNMs we choose to work with.

We then proceed to determine $\delta\omega^{(j)}$ and $\delta\tau^{(j)}$ as follows.
Using the fitting formula Eq.~(54a) of Ref.~\cite{Wagle:2021tam}, namely,
\begin{equation}
    M {\rm Re}(\sigma)_{\rm dCS} = c_{1} + c_{2} \kappa \zeta + (c_{3} + c_{4} \kappa \zeta) \, (1 - \chi_{\rm f})^{c_{5} + c_{6} \kappa \zeta},
    \label{eq:omega_dsc_eg}
\end{equation}
and similarly for the imaginary part, ${\rm Im}(\sigma)_{\rm dCS} =  - 1 / \tau_{\rm dCS}$.
Here $\kappa = 1/(16 \pi)$, $\zeta = \ell_{\rm dCS}^{4} / (M_{\rm s}^4 \kappa)$, thus
$\kappa \zeta = \gamma_{\rm dCS}$ and
where $c_{i}$ (with $i=1, \dots, 6$) are fitting coefficients which can be found in
Table II of Ref.~\cite{Wagle:2021tam},

We now expand Eq.~\eqref{eq:omega_dsc_eg} to leading orders in $\chi$ and $\gamma_{\rm dCS}$, and
gather the terms proportional to $\gamma_{\rm dCS}$.
We obtain
\begin{align}
    M \omega_{\rm dCS} =
    \left( 0.3722 + 1.1945 \gamma_{\rm dCS} \right)
    + \left(0.1861 + 5.1828 \gamma_{\rm dCS} \right) \, \chi,
    \nonumber \\
    \label{eq:omega_dcs_interm}
\end{align}
where we make use of the numerical values of the coefficients $c_{i}$.
We find (reassuringly) that the nonrotating GR part of the expression above agrees with $\omega^{(0)}$ of Ref.~\cite{Maselli:2019mjd} to 0.5\% relative error. The same estimate leads to a larger relative error ($\approx 20$\%) for the linear-in-spin coefficient (i.e., 0.1861 in comparison to 0.1258 of Ref.~\cite{Maselli:2019mjd}).
We attribute this difference to Ref.~\cite{Wagle:2021tam} having fitted
Eq.~\eqref{eq:omega_dsc_eg} to QNM data computed to linear-order in spin,
whereas~\cite{Maselli:2019mjd} fitted Eq.~\eqref{eq:kerr_expansion} to Kerr QNM
valid to all orders in spin.

We can now isolate the dCS corrections from Eq.~\eqref{eq:omega_dcs_interm} and
compare against Eq.~\eqref{eq:qnm_nongr_iso}, to find
$p_{\rm dCS} = 4$,
\begin{equation}
\delta \omega^{(0)}_{\rm dCS} = 3.1964\,, \quad \delta \omega^{(1)}_{\rm dCS} = 41.199.
\label{eq:cs_omega_coefs}
\end{equation}
We can carry the same steps for $\tau_{\rm dCS} = - 1 / {\rm Im}(\sigma)_{\rm dCS}$ and find
\begin{equation}
\delta \tau^{(0)}_{\rm dCS} = 6.3619\,, \quad \delta \tau^{(1)}_{\rm dCS} = 794.66,
\label{eq:cs_tau_coefs}
\end{equation}
which completes the set of fixed non-GR parameters in the ringdown of the \pSEOB{}
waveform model for this theory.
We remark that the alarmingly large values of
$\delta \omega_{\rm dCS}^{(1)}$ and $\delta \tau_{\rm dCS}^{(1)}$ are
compensated by the assumptions that $\gamma_{\rm dCS}$ and $\chi$
are much less than unity, which are indeed the assumptions used in Ref.~\cite{Wagle:2021tam}
to compute the QNMs.

\subsection{Effective-field-theory of general relativity}
\label{app:map_eftofgr}

The QNMs of slowly rotating BHs in both cubic and quartic EFT of GR where calculated in Ref.~\cite{Cano:2021myl}.
For the cubic EFT, we use their Eq.~(67), in the particular case of $\lame = \lamo = 1$.
We then linearize the resulting expression in $\chi$ and consider $m=2$ the harmonic.
As an outcome, we find that the fundamental axial-parity QNM is the least damped one, and it is
the one we use.
Direct comparison with Eqs.~\eqref{eq:qnm_nongr_iso} results in $p_{\rm cEFT} = 4$,
\begin{align}
    \delta \omega_{\rm cEFT}^{(0)} = - 0.5813, \quad \delta \tau_{\rm cEFT}^{(0)} = 2.6469,
    \nonumber \\
    \delta \omega_{\rm cEFT}^{(1)} = -3.8620, \quad \delta \tau_{\rm cEFT}^{(1)} = 265.12,
\end{align}
for this theory.

We proceed in the same away for the quartic EFT. Here we use Eq.~(68) (with
$\epsilon_{1} = 1$ and $\epsilon_{2} = 0$) and Eq.~(70) of
Ref.~\cite{Cano:2021myl},
In this case we find that both axial and polar modes reduce the damping time
of the fundamental QNM mode relative to GR.
Hence, we choose the axial-parity mode for this which reduction is the smallest.
This time we then find that $p_{\rm qEFT} = 6$,
\begin{align}
    \delta \omega_{\rm qEFT}^{(0)} = -0.2114, \quad \delta \tau_{\rm qEFT}^{(0)} = -0.6070,
    \nonumber \\
    \delta \omega_{\rm qEFT}^{(1)} = -1.5263, \quad \delta \tau_{\rm qEFT}^{(1)} = 171.35,
\end{align}
for this theory.
As in the case of the previous theories, the large values of some of these
coefficients are compensated by the assumptions of weak coupling and small spin
used to calculate the QNM frequencies.

\section{The estimation of  intrinsic binary parameters in General Relativity and modified theories of gravity}
\label{app:big_contour}

\begin{figure*}[t]
\includegraphics[width=0.9\textwidth]{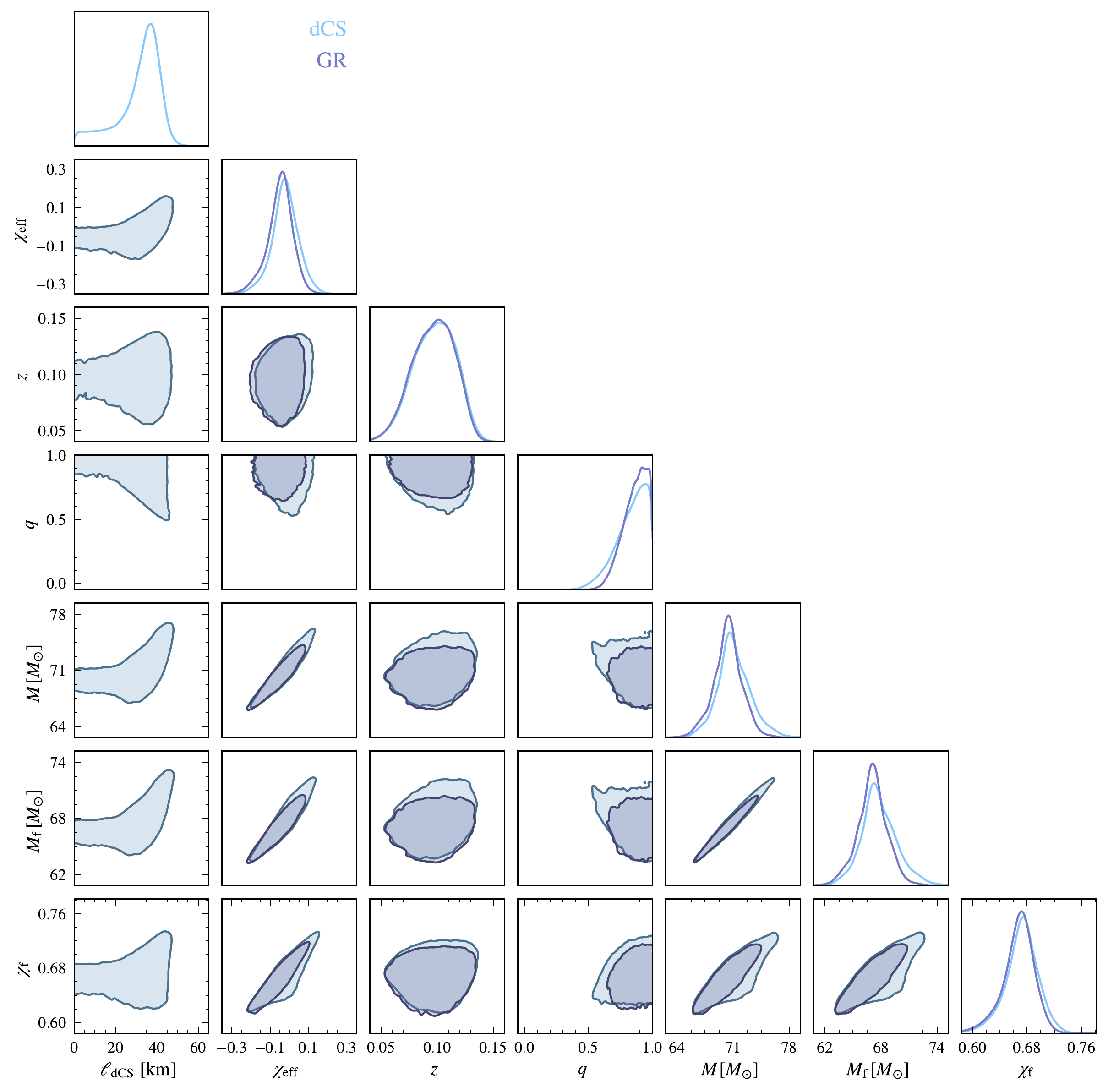}
\caption{
Corner plot showing that the inferred source binary parameters
and $\ell_{\rm dCS}$ for GW150914. We used the same waveform model \pSEOB{}
setting (purple contours) or not (blue contours) the non-GR parameters
different from zero. In the latter case, we considered dCS gravity and $N_{\rm max}=0$ as an
example.
Here, $\chi_{\rm eff}$ is the dimensionless effective-spin parameter, related to the individual spins $\chi_{i}$
and masses $m_{i}$ of each binary component as $\chi_{\rm eff} \equiv (m_1 \chi_1 + m_2 \chi_2) / (m_1 + m_2)$.
All contours correspond to 90\% credible levels.
We see that the addition of the non-GR parameters does not introduce biases in the
inference of the source parameters.
We found the same qualitative behavior in the posteriors distributions of the
source binary parameters for the other modified gravity theories studied in the
main text.
The same conclusions apply for the other GW event studied here, GW200129.
}
\label{fig:corner_plot_all}
\end{figure*}

We show in Fig.~\ref{fig:corner_plot_all} a corner plot for all the intrinsic
binary parameters from our parameter-estimation study of GW150914, using the
\pSEOB{} waveform models for GR and dCS. We also included $\ell_{\rm dCS}$ in
the latter case.
We see that the medians of the posterior distributions are not affected
substantially by the inclusion of the non-GR parameters.
This is particularly important for the $\Mf$ and $\chi_{\rm f}$ parameters, since
the ParSpec framework assumes that the non-GR theory induces small deviations from GR.
Indeed, since \pSEOB{} introduces only minimal modifications to the plunge-merger
and because the remnant BH parameters are estimated according to GR
predictions using the binary's component masses and spins, the fact that only small
biases are introduced on $\Mf$ and $\chi_{\rm f}$ is, to some extend, expected.
We obtain qualitative similar results for GW200129 and the other modified
gravity theories considered in our work.
We also remark that the fact that the posterior on $\chi_{\rm f}$ has most
support around $\approx 0.7$ justifies our use of the fitting coefficients in
the ParSpec formulas of Ref.~\cite{Maselli:2019mjd}.

\bibliography{paper_alt_theory_bounds}

\end{document}